\UseRawInputEncoding
\documentclass{article}
\usepackage{color}   %May be necessary if you want to color links
\usepackage{hyperref}
\hypersetup{
    colorlinks=false, %set true if you want colored links
    linktoc=all,     %set to all if you want both sections and subsections linked
    linkcolor=blue,  %choose some color if you want links to stand out
}
\usepackage{caption}
\usepackage{subcaption}

\usepackage[round, sort, numbers]{natbib}
\usepackage[ruled,vlined]{algorithm2e}
\usepackage[english]{babel}
\usepackage{amsmath}
\usepackage{verbatim}
\DeclareMathOperator\erf{erf}
\usepackage[autostyle]{csquotes}
\usepackage{amsfonts}
\usepackage{amssymb}
\usepackage{float}
\makeatletter
 \usepackage{url}

\makeatother

\usepackage{geometry}

\usepackage{graphicx}
\usepackage{multicol}
\usepackage[version=4]{mhchem}
\usepackage[table,xcdraw]{xcolor}
\usepackage{float}
\title{Strong Connectivity in Real Directed Networks}

\author{Niall Rodgers, Peter Ti\v{n}o and Samuel Johnson }

\date{\today}% It is always \today, today,
\begin{document}
\setcitestyle{round}
\maketitle
\begin{abstract}
%    Strong Connectivity is of vital importance to many real-world directed networks. We show that it can be predicted whether a real-world network is strongly connected from only the average degree and a measure of global directionality and hierarchical ordering called Trophic Incoherence, which can be applied to any directed network. We demonstrate that strong connectivity is driven by the edges which break this hierarchical ordering and derive the critical point, using percolation theory, where the giant strongly connected component forms using these edges. We confirm our results by studying a range of real-world networks including food webs, neural networks, trade networks and more. We show that the connectivity structure can be disrupted with minimal effort by a targeted attack on these backwards edges. We demonstrate through the example dynamics including the SIS model, Majority Vote, Kuramoto Oscillators and the Voter Model the effects of a targeted attack on these edges and how understanding the global directionality can aid our intuition around dynamical processes.       
    
    % Suggested edited abstract:
In many real, directed networks, the strongly connected component of nodes which are mutually reachable is very small. This does not fit with current theory, based on random graphs, according to which strong connectivity depends on mean degree and degree-degree correlations. And it has important implications for other properties of real networks and the dynamical behaviour of many complex systems.
We find that strong connectivity depends crucially on the extent to which the network has an overall direction or hierarchical ordering -- a property measured by trophic coherence.  Using percolation theory, we find the critical point separating weakly and strongly connected regimes, and confirm our results on many real-world networks, including ecological, neural, trade and social networks. We show that the connectivity structure can be disrupted with minimal effort by a targeted attack on edges which run counter to the overall direction. And we illustrate with example dynamics -- the SIS model, majority vote, Kuramoto oscillators and the voter model -- how a small number of edge deletions can utterly change dynamical processes in a wide range of systems.

\end{abstract}

\section{Introduction}

Understanding the connectivity structure of a directed network is vitally important in many different contexts. Can every node be reached in a communications network or one way street grid? How will a disease spread or will a dynamical system be stable and resilient to perturbation? Whether a network will be connected has been well studied in the case of undirected networks through percolation theory, however this is less well understood in directed networks and hence real-world systems which are often directed \cite{Li2021PercolationApplication,Johnson2020DigraphsSystems}. We demonstrate through understanding the global directionality and hierarchical organisation of directed networks through a method known as Trophic Analysis \cite{MacKay2020HowNetwork} that it is possible to construct a phase diagram which predicts if real networks are strongly connected using only the average degree and the incoherence parameter which measures the global directionality going beyond previous understanding based on directed random graphs \cite{Boguna2005GeneralizedNetworks}. The notion of global directionality provided by trophic analysis enables us to talk meaningfully of ``forward" and ``backwards" edges and gives us an insight into the directed network which is not possible without this. It is the backward edges that break the overall hierarchical structure. This hierarchical structure can be found in almost all real-world networks, not just networks with obvious hierarchy such as food webs where the hierarchy is number of steps from the nodes of zero in-degree such as plants. Global directionality, Trophic Incoherence, has been linked to network non-normality \cite{MacKay2020HowNetwork} which has been shown to be ubiquitous in real directed systems \cite{Asllani2018StructureNetworks,Duan2022NetworkSystems}. We use the insight that the strong connectivity is driven by edges which break the hierarchical ordering to apply percolation theory \cite{Coupette2022ExactlyProblems} to these ``backwards'' edges and an analytical estimate of the number of such edges derived from the global directionality to analyse the connectivity structure. This provides an insight beyond degree \cite{Boguna2005GeneralizedNetworks} and explains why, even if they have high degree, highly structured networks like food-webs are not strongly connected. We extend our understanding of strong connectivity to real networks beyond previous results on directed random graphs \cite{Boguna2005GeneralizedNetworks} which do not capture the complex structures of real-world systems. We demonstrate the role the ``backwards'' edges have in controlling the strong connectivity by conducting a targeted attack on these edges. This removes the strongly connected component while maintaining the weak connectivity. We show the vital role that strong connectivity plays in dynamics by comparing the spread of an infection using SIS dynamics; synchronisation of coupled Kuramoto Oscillators and how a new state establishes itself in the Majority Vote and Voter Models before and after the targeted attack. These dynamics demonstrate the role of hierarchy in the dynamics as the global directionality and ``backwards'' edges drive feedback and how their removal creates an asymmetry the ability of nodes to interact with each other dependent on their position in the hierarchy.

\section{Background}
Trophic Analysis is a technique which is used to calculate the global directionality and the hierarchy in directed complex networks \cite{MacKay2020HowNetwork}. Complex networks are graphs which represent real-world systems. Graphs are sets of vertices (nodes) and edges (links) which represent connections between elements in the system. Graphs are topological objects as they do not need to have a distance scale they merely represent that elements are connected. A directed graph is one in which the connections between elements go in only one direction. This is very common in real-world systems \cite{Johnson2020DigraphsSystems} which can be intrinsically directional like a prey-predator food web interaction or following a profile on social media. In complex networks it is common to represent a graph via an adjacency matrix. For a graph consisting of $N$ nodes the adjacency matrix, $A$ is defined such that  \begin{equation} \label{Adj_eq}
A_{ij}=
    \begin{cases}
     1  \text{ \quad if there exists an edge } i \to j \\ 
     0  \text{ \quad  otherwise }
    \end{cases}.
\end{equation}   
As a result, the topology of the graph can be represented by the non-zero entries of this matrix. This form is preferred for studying complex networks as it is convenient for computer simulations, defining dynamical systems on the network and  network properties can easily be accessed from the properties of this matrix. In a directed graph this matrix is not necessarily symmetric,  $A_{ij}\neq A_{ji}$, since the interactions are only in one direction. Undirected graphs always have symmetric adjacency matrices. Adjacency matrices can also be weighted to capture the strength of an interaction, however for simplicity we focus on the unweighted case. An undirected graph is connected if and only if for any pair of distinct nodes there exists a path connecting them. In directed graphs the notion of connectivity is more complex. A directed graph is weakly connected if there is a path between all pairs of vertices when edge direction is ignored. A digraph is strongly connected if there is a path between all pairs of vertices when the edge direction is respected. In real-world directed complex networks is it very common that they are weakly connected but may not be strongly connected. Another important aspect to define is the degree of a node. This is the number of connections a node has to the rest of the network. In a directed graph the in-degree of a node $i$ is the number of connections which point to it $k_{i}^{in} = \sum_{j} A_{ji}$ and the out-degree is the number of out neighbours it has, $k_{i}^{out} = \sum_{j} A_{ij}$. If the adjacency matrix is transposed the in and out degrees swap. \footnote{This is a point to take care with as many authors have opposite conventions for defining the adjacency matrix.}

\subsection{Trophic Analysis}

Trophic Analysis first arose in the study of food webs \cite{Johnson2014TrophicStability} where the hierarchical organisation of the network was proposed as a solution to May's paradox \cite{May1972WillStable}, surrounding the stability of food webs. The name arises from the Trophic level of a species in ecology \cite{Levine1980SeveralWebs}. This definition relies on the existence of basal nodes, nodes with in-degree zero, however it was redefined in \cite{MacKay2020HowNetwork} to remove this constraint and make it applicable to any directed network. We follow the updated convention \cite{MacKay2020HowNetwork} but the reader should note that most previous work used the previous convention \cite{Johnson2014TrophicStability}.  Trophic Analysis has been used to study the structure in food webs \cite{Klaise2017TheWebs}; spreading processes such as epidemics or signals in neural networks \cite{Klaise2016FromProcesses}; resilience of infrastructure networks \cite{Pagani2019ResilienceNetworks,Pagani2020QuantifyingNetworks} and control of organisations \cite{Pilgrim2020OrganisationalAnarchy}.

Trophic Analysis is composed of two parts: the node level information, Trophic Level, and the global information, Trophic Incoherence \cite{MacKay2020HowNetwork}. Trophic level gives a measure of where a node sits in the hierarchy of a directed network. For example, in a food web plants would be the low trophic level nodes and carnivores the high trophic level nodes, as energy flows up the food web from low to high trophic level. This can however be generalised to any directed network. Trophic level,  is calculated by solving the $N \times N $ matrix equation given by \begin{equation} 
    \Lambda h = v.
    \label{eq_h}
\end{equation}
Where $h$ is the vector of trophic levels for each node and $v$ is the vector imbalance of the in and out degree of each node where each element is defined as $ v_i = k_{i}^{in} - k_{i}^{out}$. $\Lambda$ is the Laplacian matrix:
%Which is 
\begin{equation}
    \Lambda = diag(u) - A - A^{T}.
\end{equation}

Which depends on the sum of the in and out degrees of each node, $ u_i = k^{in}_i + k^{out}_i$, the adjacency matrix, $A$, of the graph and its transpose, $A^{T}$. These are all quantities which can be simply evaluated from the adjacency matrix and the equation for trophic level can be solved iteratively as the Laplacian matrix may have a zero eigenvalue. Equation \ref{eq_h} is invariant under the addition of a constant vector to $h$ since the Laplacian vanishes when it acts on a constant vector so the solution of the equation can be modified to adhere to the convention that the lowest level node takes the value 0. However, this is just a convention and not a requirement.

Trophic Incoherence measures the global directionality of the network \cite{MacKay2020HowNetwork}. It provides a measure of distribution of level differences across the edges in the system. If the network is fully coherent,  the edges only connect to nodes exactly one level above them and the network is perfectly hierarchical and globally directed. If the network is incoherent then the edges connect without respect to the levels and there is no global directionality. This is quantified via the trophic incoherence  parameter $F$  which is defined as \begin{equation}
    F = \frac{\sum_{ij}A_{ij}(h_{j} -h_{i}-1)^2}{\sum_{ij}A_{ij}}.
    \label{eq_F}
\end{equation}   
This equation measures, averaged over the system, the square of the deviation in level difference of destination to source vertex from 1 across the edges of the graph. This equation is bound between 0 and 1 \cite{MacKay2020HowNetwork}. Networks with $F=0$ are perfectly coherent, they have distinct integer levels in which all nodes are placed and are acyclic. When $F=1$ every node has the same level and the network has no hierarchy. Examples of $F=1$ networks are directed cycles. Networks which have $F=1$ are perfectly balanced so hard to numerically generate through a random process, for example random graph models such as Erd\H{o}s-R\'{e}nyi model  \cite{Erdos1959OnI,Johnson2017LooplessnessCoherence} lead to networks where $F$ is around 0.95 although this may vary depending on sparsity. The trophic levels can be thought of as the set of levels, $h$ which minimise $F$ \cite{MacKay2020HowNetwork} with minimising equation \ref{eq_F}  leading to to equation \ref{eq_h}. It is also possible to equivalently  speak in terms of Trophic Coherence which can be defined as $1-F$. When we say top of the hierarchy we mean the nodes with high trophic level and when we use the phrase bottom of the hierarchy we mean nodes of low trophic level.

\subsubsection{Generated Networks}
 When we require generated networks to better sample the full range of Trophic Incoherence and degree we the use the same variant of the generalised preferential preying modelled as \cite{Rodgers2022NetworkNetworks} which was based on work from \cite{Klaise2016FromProcesses}. This model allows the Trophic Incoherence of a generated network to be roughly controlled although not to a fully precise level of detail. This is done by a taking an initial network structure and adding edges with a probability which is proportional to the level difference between the nodes in a way which is controlled by a temperature parameter. This is defined as \begin{equation}
        P_{ij} \propto \exp{\left[ -  \frac{(\tilde{h}_{j} - \tilde{h}_{i} -1 )^2}{2T_{\text{Gen}}}\right]}.  
    \end{equation}    

Where $\tilde{h}_{i}$ which is the temporary trophic level assigned during the generation. $T_{Gen}$ is the generation temperature used to control the incoherence. At high $T_{Gen}$ edges are added without respect to the level structure so it produces an incoherent network similar to a random graph, while at low temperature the edges are only added when the level difference is near one so produces a very coherent network. Details of how to efficiently sample the possible edges and generate networks in this way can be found in \cite{Rodgers2022NetworkNetworks}.

\section{Results}

\subsection{Fraction of Edges going against the Hierarchy}

It is possible to analytically estimate the number of edges that go ``backwards'' - against the hierarchy. These edges are important as they determine the strong connectivity of the network because they are needed to induce a path back down the hierarchy. This fraction of edges is useful in further calculation of strong connectivity. 

The first part of this derivation is to assume that the edges follow an approximately Gaussian distribution in trophic level differences \cite{Johnson2017LooplessnessCoherence}, where the mean is the mean level difference, $\overline{z}$ and the standard deviation is given by $\overline{z}\eta$. The mean level difference can be computed from the Trophic Incoherence, \begin{equation}
    \overline{z} = 1-F,
\end{equation}
with the derivation given in \cite{MacKay2020HowNetwork}. The parameter $\eta$ can also be expressed in terms of $F$ \cite{MacKay2020HowNetwork} given as \begin{equation}
    \eta = \sqrt{\frac{F}{1-F}}.
\end{equation}

Assuming that the edge level differences, $x_{ij} = h_j - h_i$, follow a Gaussian distribution leads to the probability distribution, \begin{equation}
    p(x_{ij}) = \frac{1}{\overline{z}\eta\sqrt{2\pi}}\exp{\left[-\frac{1}{2}\left(  \frac{x_{ij} - \overline{z}}{\overline{z}\eta} \right)^2\right]}.
\end{equation}  

The fraction of edges which do not go in the same direction as the hierarchy is the integral of this distribution from negative infinity to 0. The cumulative distribution of a Gaussian is well known and the result can be written in terms of the error function as \begin{equation} \label{back_edge_eq}
    \overline{\beta (F)} =  \frac{1}{2} \left[ 1 +  \erf{\left(-\frac{1}{\sqrt{2}}\sqrt{\frac{1-F}{F}} \right)}\right].
    \end{equation}
    
This equation can be understood by looking at the limiting cases where $F$ equals 1 or 0. When $F=1$ the error function goes to zero and then half the edges go against the ``hierarchy" which makes sense as every node has the same level hence an equal likelihood of going forwards or backwards. When $F=0$ the error function goes to negative 1 so the expression cancels and no edges go backwards, which makes sense as the network is fully coherent.

\begin{figure}[H]
      		\centering
            \includegraphics[width=0.8\linewidth]{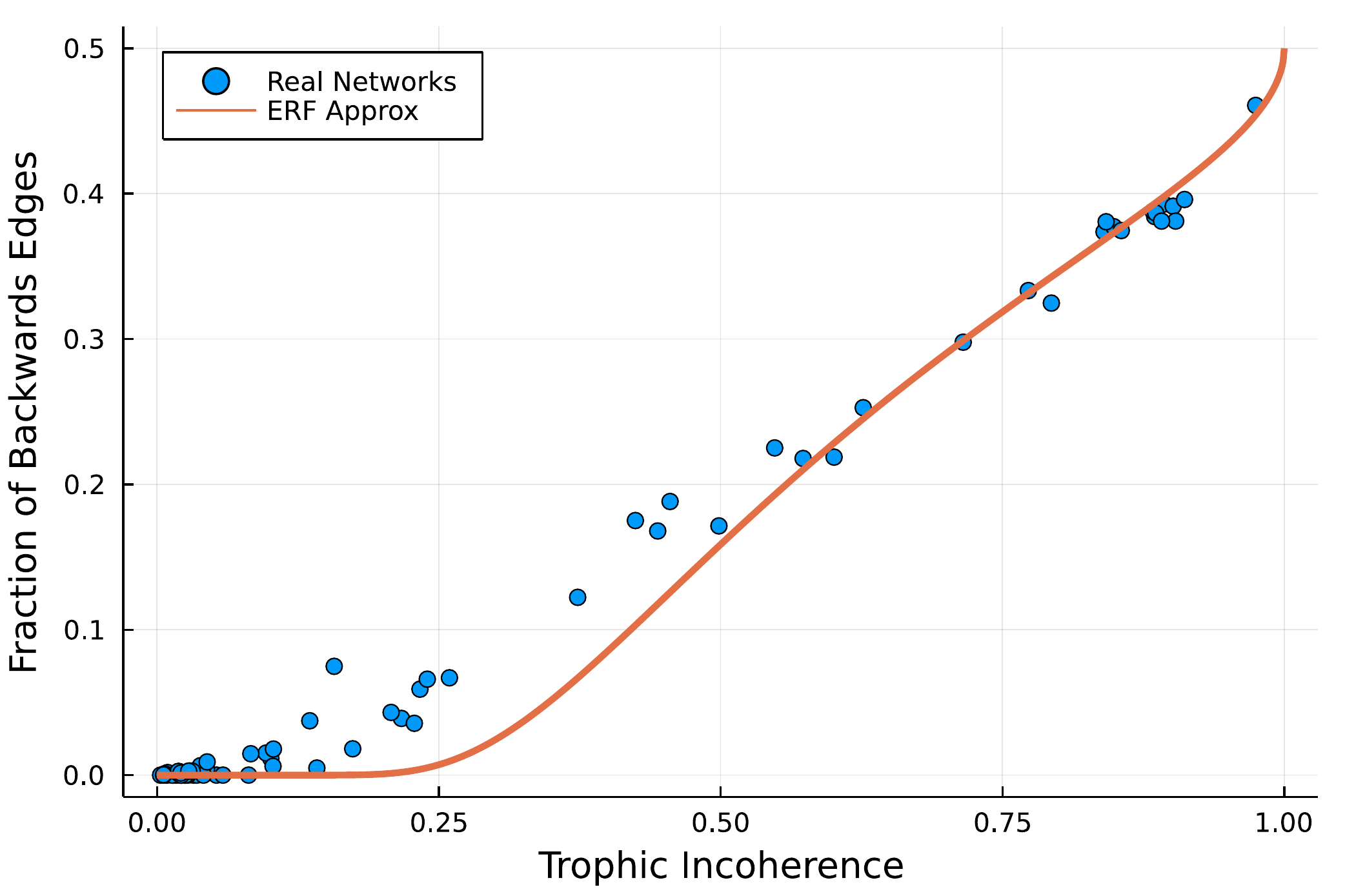}
        	\caption{ Number of backwards edges in real networks \cite{DataSamJohnson} and analytical prediction}

        \label{fig:error_function_real}
        \end{figure}

This prediction holds well in real networks shown in figure  \ref{fig:error_function_real} with some small deviation which is not surprising due to the variability of real networks and the assumption that distribution of the edge differences is Gaussian. All the real networks used in this paper and the sources are linked online at \cite{DataSamJohnson}. This data set includes metabolic networks, neural networks, trade networks, food webs and social networks. The number of backwards edges could also provide a rough estimate of the upper bound on the size of a feedback arc set, the number of edges which need to be removed to make the graph acyclic \cite{Sun2017BreakingHierarchies}. The link between hierarchy breaking edges and a heuristic to approximate the cycles has been made before \cite{Sun2017BreakingHierarchies} with different measures of the hierarchical ordering, including PageRank \cite{Google1999}, however these lack an analytical estimate of the expected number of the backwards edges and hence the link to the strong connectivity which follows. 

% This analytical result is useful and it helps our intuition that the edges which go backwards are important for the growth of strong connectivity. It is possible to adapt the derivation of the size of the connected component in a directed ordered network undergoing explosive percolation found in \cite{Waagen2017ExplosiveActivity} and it leads to an estimate of the strongly connected component in growing exponentially in $\overline{\beta (F)}\langle k \rangle$ which recovers the results found in the paper but provides an underestimate of the growth rate and misses the transition point. However the transition point can be found by adapting the interpretation found in \cite{Coupette2022ExactlyProblems}.   

\subsection{Derivation of Strong Connectivity Critical Point and Phase Diagram}

It is possible to derive an estimate of the percolation transition threshold for strong connectivity in directed networks using the insight gained from the hierarchical structure. This can be done by observing that if the nodes in a network are ordered in some way, then the edges which break that ordering by going `backwards' are the important edges for strong connectivity. Adding more edges in the forward direction will not make the network strongly connected if it is very strictly hierarchical as they do not provide a way to move back down the ordering towards the bottom of the network. In this way the growth of the strongly connected component in a directed network can be thought of as a percolation process on the backwards edges, where the backwards edges connect the layers of the network. This makes it possible to move back down the hierarchy therefore creating a giant strongly connected component.

This can be expressed using the framework for solving percolation problems set out in \cite{Coupette2022ExactlyProblems}. This framework decomposes the percolation process to transitions between l-step neighbourhoods which in our case can be thought of as steps down the hierarchy. We assume a weakly connected network when the backwards edges are removed. Which is a reasonable assumption for real networks as the fraction of backwards edges is usually small and the networks are dense enough so that removal of the backwards edges does not result in the network being disconnected. It is however possible to find counter examples to this where the network does become disconnected if the backwards edges are removed (see Appendix \ref{Weak_connectivty}). This however only occurs when the backwards edges are calculated once and trophic level is not recalculated after each removal. This is proven in Appendix \ref{Weak_connectivty} where we also show examples of the maintenance of weak connectivity in real networks justifying this assumption. We wish to find the percolation threshold for the network to be at least weakly connected by only backwards edges and hence have a giant strongly connected component. 
\paragraph{}

We define a directed subgraph, $G(V,E_B)$, made up only of backwards edges, where the trophic level difference is less than or equal to 0, of the larger graph $H(V,E)$ containing all the edges from which the trophic levels are calculated. Following the steps laid in out in \cite{Coupette2022ExactlyProblems}, we introduce the $l$-neighbourhood of a vertex, $y$. This is recursively defined as 

\begin{equation}
    \mathcal{N}_l(y) = \bigcup_{X \in V(\mathcal{N}_{l-1}(X)) } \mathcal{N}_1(X).
\end{equation}
Where $V(\mathcal{N}_{l-1}(X)) $ is the set of all of the vertices within the neighbourhood $\mathcal{N}_{l-1}(X)$. This neighbourhood can be thought of as the nodes reachable in within $l$ steps from vertex $y$, illustrated in  more detail in \cite{Coupette2022ExactlyProblems}. The percolation transition can then be understood by analysing the surfaces of these neighbourhoods, which can be defined as the vertex sets,
\begin{equation}
    V_l := V({\mathcal{N}}_l(y) )\setminus V({\mathcal{N}}_{l-1}(y)). 
\end{equation}
These are the nodes which lie exactly $l$ steps from the origin vertex. Where $y$ is an origin which can in general be any vertex but in the case of backwards connectivity we choose the vertex with the highest trophic level. Following the work of \cite{Coupette2022ExactlyProblems} the system is above the percolation threshold if
\begin{equation} \label{perc_threshold}
    \lim_{l\to\infty} \mathbb{E}[o(V_l)] > 0.
\end{equation} 
Where $\mathbb{E}[x]$ is the expectation value of $x$ and $o(V_l)$ is the number of connected nodes in the surface $l$. The expectation values are taken using draws from the ``coherence ensemble'', the set of all unweighted directed networks of fixed trophic coherence, size and degree distribution, used in \cite{Johnson2017LooplessnessCoherence,MacKay2020HowNetwork}. More detail can be found in \cite{Coupette2022ExactlyProblems} but equation \ref{perc_threshold} can be understood as that there is a giant connected component if the expectation value of a node being connected is greater than zero as the surface size extends to infinity. This is analogous to the probability of a branching process dying out as the number of steps tend to infinity.  Our assumption is that in the network of backwards edges the expected number of connected nodes in a surface is simply the number of connected nodes in the previous surface multiplied by the average number of a backwards connections. This is \begin{equation}
    \mathbb{E}[o(V_{l+1})] = \beta\langle k \rangle \mathbb{E}[o(V_{l})],
\end{equation}    
where $\langle k \rangle$ is the mean total degree and $\beta$ is the fraction of edges which go backwards. 
This equation can then be solved iteratively assuming that $\mathbb{E}[o(V_{0})] =C$, where $C$ is some finite constant representing the number of nodes at the top of the hierarchy. Leading to  \begin{equation}
    \mathbb{E}[o(V_{l})] = (\beta\langle k \rangle)^l C. 
\end{equation}

Taking the limit $l$ goes to infinity leads to the result,

\begin{equation} 
    \lim_{l\to\infty} \mathbb{E}[o(V_l)] = \begin{cases} \infty \quad \text{if} \quad \beta\langle k \rangle > 1 \\
    C \quad \text{if} \quad\beta\langle k \rangle = 1\\
    0 \quad \text{if} \quad \beta\langle k \rangle < 1
    
    \end{cases}.
    \end{equation} 

This means that we expect a giant connected component when $\beta\langle k \rangle > 1 $, which means on average each node has at least one backwards connection. This can also be written directly as a function of trophic incoherence using the expected value of $\beta$, \begin{equation}\label{eq:perc_F}
    \frac{\langle k \rangle}{2} \left[ 1 +  \erf{\left(-\frac{1}{\sqrt{2}}\sqrt{\frac{1-F}{F}} \right)}\right] > 1.
\end{equation}

\begin{figure}[H]
     \centering
\begin{subfigure}[b]{0.48\textwidth}
      		\centering
            \includegraphics[width=\textwidth]{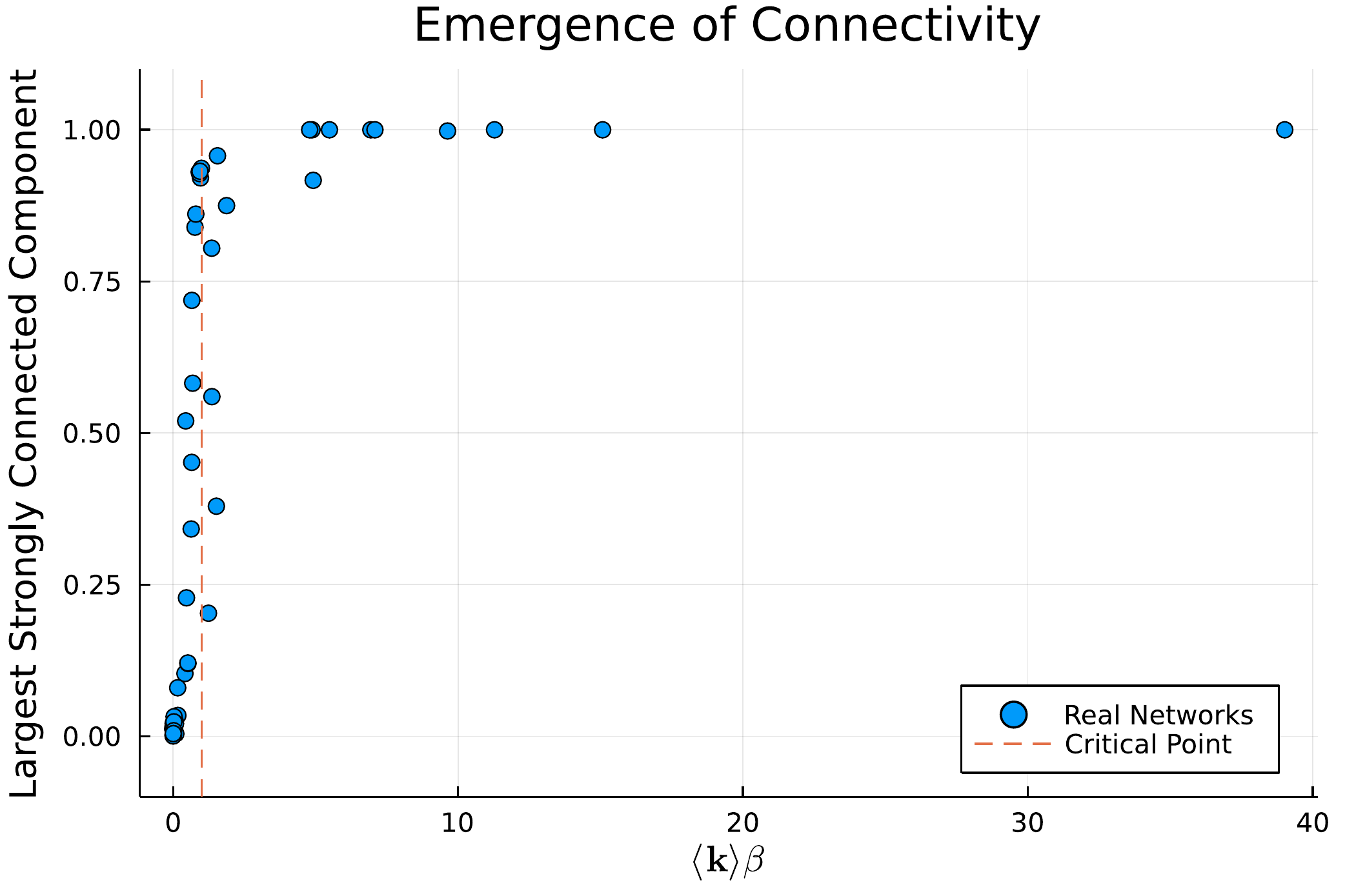}
        	\caption{ Full Range of Real Networks  
        	}
        	
        \label{fig:Connectivty_real}
        \end{subfigure}  
\begin{subfigure}[b]{0.48\textwidth}
      		\centering
            \includegraphics[width=\textwidth]{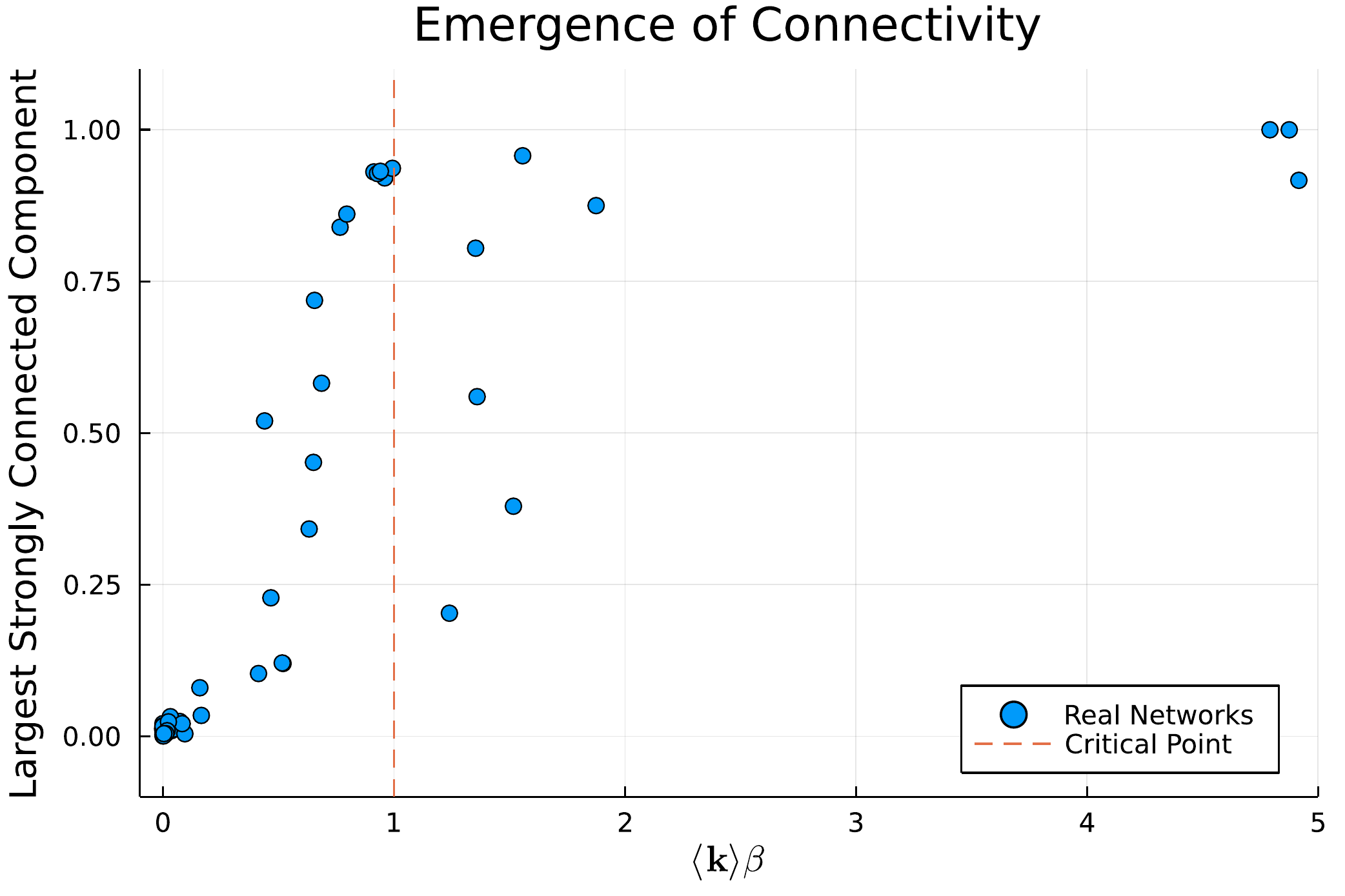}
        	\caption{ Closer View of Critical Point}
        	
        \label{fig:Connectivty_zoom_real}
        \end{subfigure} 
        \caption{ Predicted Critical point and emergence of giant strongly connected component in real networks \cite{DataSamJohnson}}
        
\end{figure}

This estimate, which uses very little information about the network structure, works well for real networks, figures \ref{fig:Connectivty_real} and \ref{fig:Connectivty_zoom_real}. This is a very interesting result which shows how the understanding of hierarchy can allow insights into the connectivity of directed networks. This result relies upon the ability to calculate the hierarchy for any directed network, the realisation that the backwards nodes shape connectivity and that their number can be linked to the global directionality and analytically estimated.
Other measures of hierarchy would allow the number of ``backwards'' edges to be enumerated numerically but lack the link to global directionality which gives the intuition behind these results. 
%Certain real networks may have specific constraints which induce a hierarchy. For instance, space in a transport network or social status in a social one. But the benefit of trophic levels, as defined here, is that they emerge 

\begin{figure}[H]\label{fig:coonectivity_predict}
     \centering
\begin{subfigure}[b]{0.45\textwidth}
      	\centering
            \includegraphics[width=\linewidth]{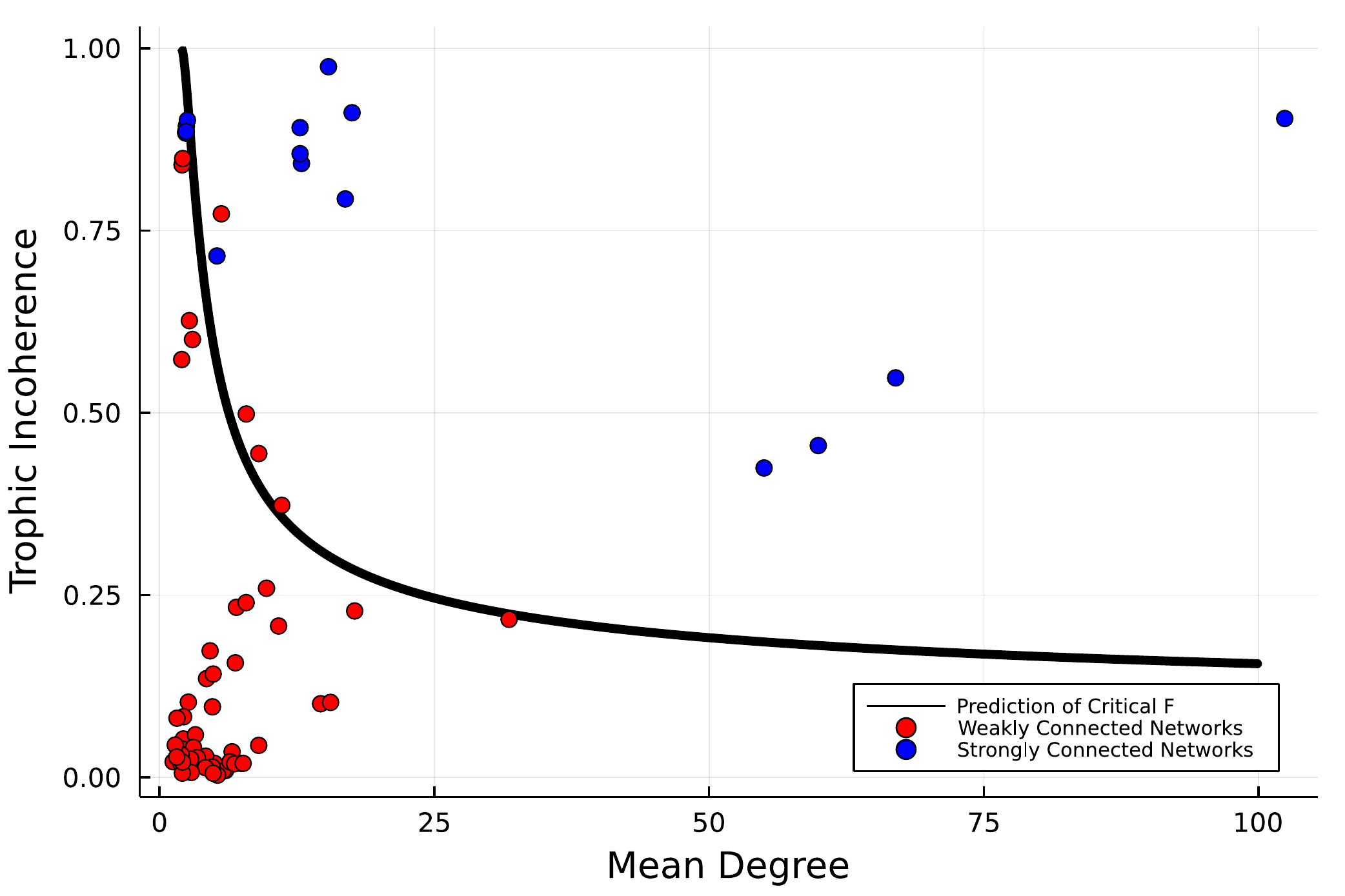}
        	\caption{ Real Networks from \cite{DataSamJohnson}.    
        	}
        	
        \label{fig:Connectivty_predict_real}
        \end{subfigure}  
\begin{subfigure}[b]{0.45\textwidth}
      	\centering
            \includegraphics[width=\linewidth]{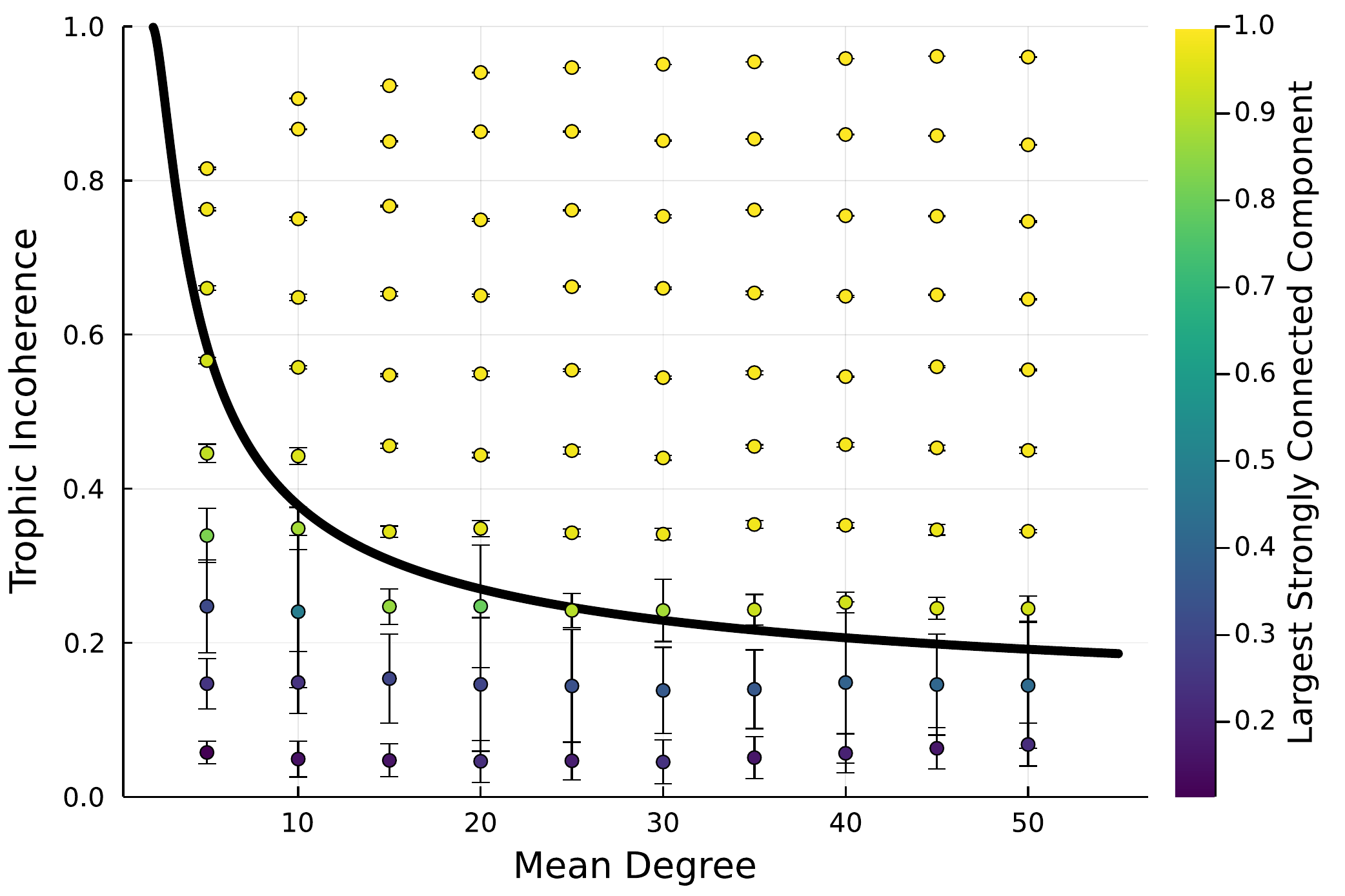}
        	\caption{ 1000 generated networks with $N=500$ generated as in \cite{Rodgers2022NetworkNetworks} with varying degrees and binned by Trophic level. Error bars are the standard deviation of the averaging of the incoherence. 
        	}
        	
        \label{fig:Connectivty_predict_generated}
        \end{subfigure} 
        \caption{Prediction of Strong Connectivity using the mean degree and Trophic Incoherence based on the critical F derived via the backwards connectivity. }
        
\end{figure}

The equation for the percolation threshold, under the Gaussian edge difference assumption, in terms of F, equation \ref{eq:perc_F}, can be expressed as an equation for the critical incoherence. This can be written as \begin{equation} \label{eq:Critical_F}
    F_{c} = \left[ 1 + 2\left(\erf^{-1}{\left(\frac{2}{\langle k \rangle} -1\right)}\right)^2\right]^{-1}.
\end{equation}  
For a fixed $\langle k \rangle $ if $F$ is greater than the predicted value then we expect the network to be strongly connected. This allows the strong connectivity of a real network to be predicted based only on $F$ and the average degree. The accuracy of this prediction for real networks is demonstrated in figure \ref{fig:Connectivty_predict_real}. This is on the whole quite accurate where we are assuming the network is strongly connected if 90\% of the nodes are in the largest strongly connected component. The only regions where the prediction is less good is close to the boundary however this is not surprising as we are not taking account of any finite size affects or potentially heterogeneous degree distributions and in particular how the backwards edges are distributed. These results are broken down by network type in the supplementary information. We can give further insight into the accuracy of the prediction by using numerically generated networks to better sample the parameter space and verify the results in a larger region. This is shown in figure \ref{fig:Connectivty_predict_generated}. We take 1000 networks where $N=500$ generated as in \cite{Rodgers2022NetworkNetworks} where each node has at least in-degree 1 which makes the trophic level impossible to calculate in the previous definition \cite{Johnson2014TrophicStability} and then bin them by Trophic Incoherence and average the size of the strongly connected component. This result agrees well with the analytical prediction with the networks well above the boundary being strongly connected and a large component forming in the networks around the boundary as expected. These results which hold for both real and generated networks, given the assumption of Gaussian edge differences and the small amount of degree information used, give a good insight into how global directionality determines strong connectivity. Even for very large degree a network is still unlikely to be strongly connected if F is low enough which demonstrates that more than just degree is needed for determining the connectivity of a directed network. Why some real networks lie directly on the transition line and properties of networks at this point may be a possible outlet for future work. 

%\begin{figure}[H]
      		%\centering
            %\includegraphics[width=0.8\linewidth]{Critical_F_connectivty.pdf}
        	%\caption{ Prediction of Strong Connectivity using the mean degree and %Trophic Incoherence based on the critical F derived via the backwards connectivity. Networks from \cite{DataSamJohnson}.    
        	%}
        	
        %\label{fig:Connectivty_predict}
        %\end{figure}    

\begin{figure}[H]
      		\centering
            \includegraphics[width=0.8\linewidth]{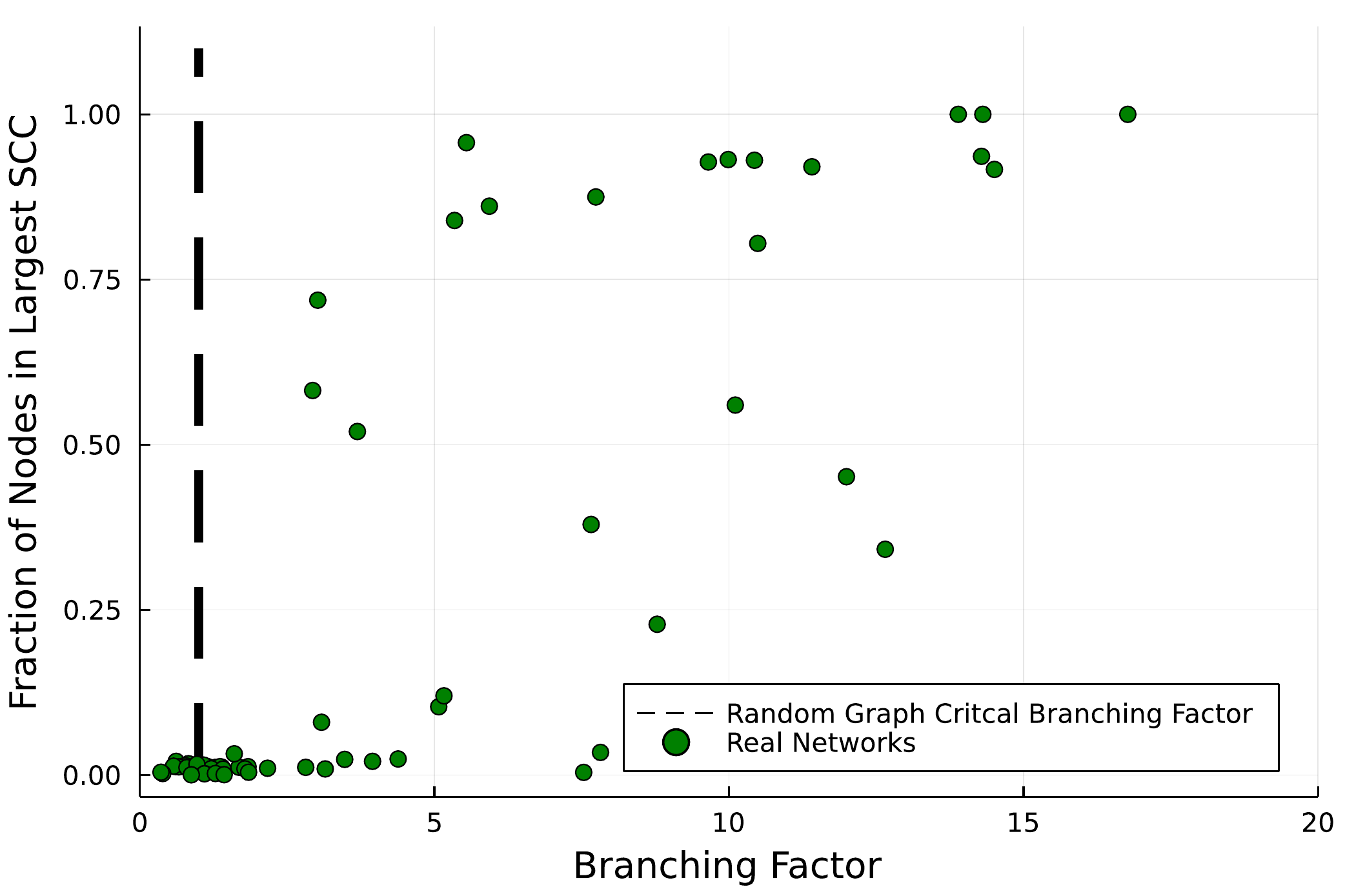}
        	\caption{ Prediction of Strong Connectivity using the Branching Factor for Real Networks \cite{DataSamJohnson}.    
        	}
        	
        \label{fig:Connectivty_branching}
        \end{figure}

The value of this analysis can be highlighted by comparing to the results obtained by taking this real network data set and trying to predict the strong connectivity without using the hierarchical structure. This can be roughly estimated using results from \cite{Boguna2005GeneralizedNetworks,Johnson2017LooplessnessCoherence} where one wold expect the strongly connected component to grow very quickly and the percolation to occur when the branching factor is greater than 1, \begin{equation}
    \frac{\langle k^{\text{in}}k^{\text{out}} \rangle }{\langle k \rangle } > 1.
\end{equation} 
This is demonstrated in figure \ref{fig:Connectivty_branching} which shows how many networks of very high branching factor are not strongly connected using this analysis. The figure represents a closer look at the critical point and networks of very high branching factor are not shown. The full result can be found in the supplementary information. This demonstrates how the directional organisation is a vital part of the connectivity structure of real networks. Understanding the interplay between global directionality, Trophic Incoherence, and ordering, Trophic Level, provides an intuition greater than each individual notion can.

\subsection{Targeted Attacks on Backwards Edges}

\begin{figure}[H]\label{fig:back_edges_attack}
     \centering
\begin{subfigure}[b]{0.48\textwidth}
      		\centering
            \includegraphics[width=\linewidth]{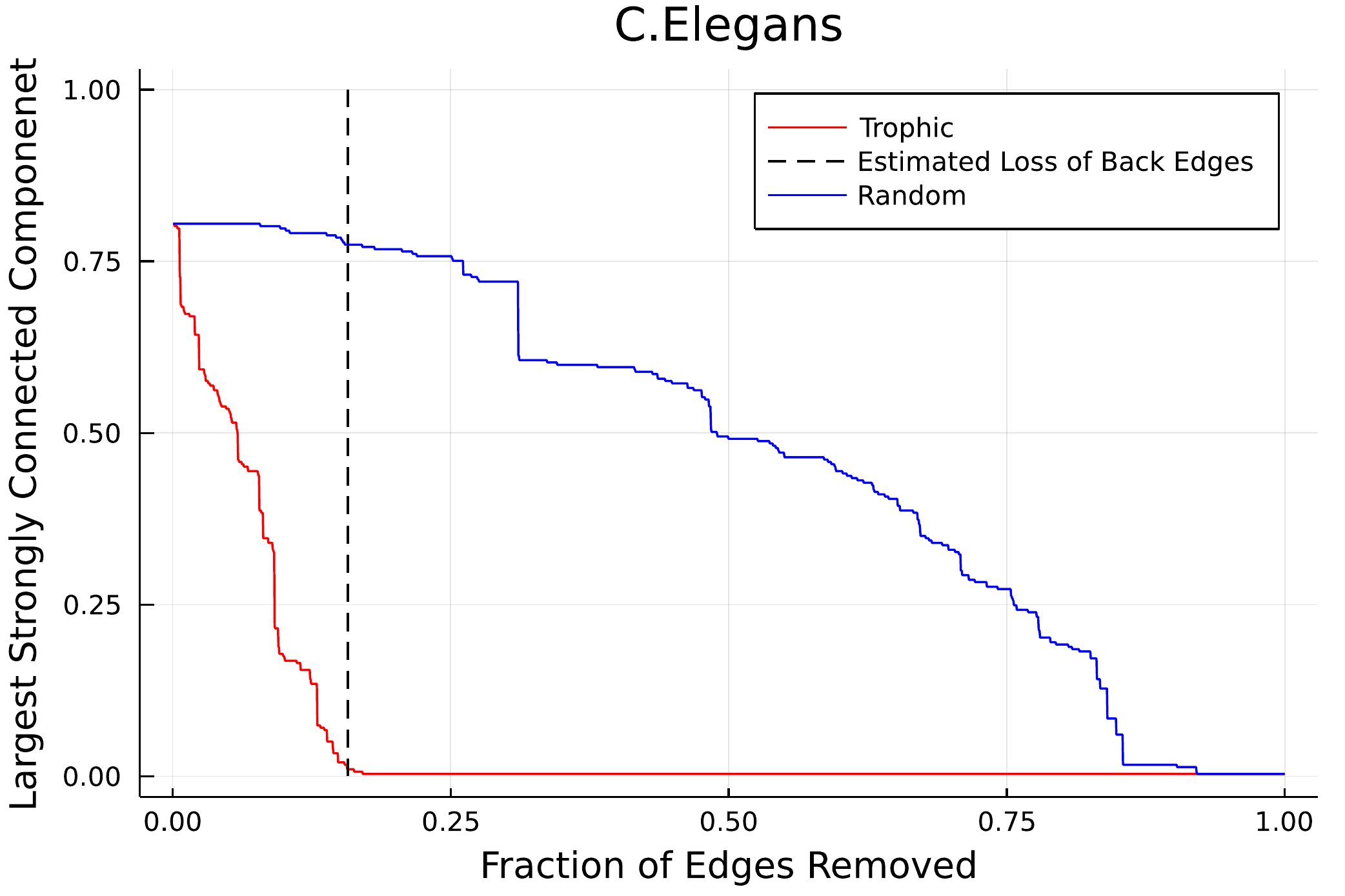}
        	\caption{  \textit{C.Elegans} connectome.  
        	}
        	
        \label{fig:Target_C_elegans}
        \end{subfigure}  
\begin{subfigure}[b]{0.48\textwidth}
      	\centering
            \includegraphics[width=\linewidth]{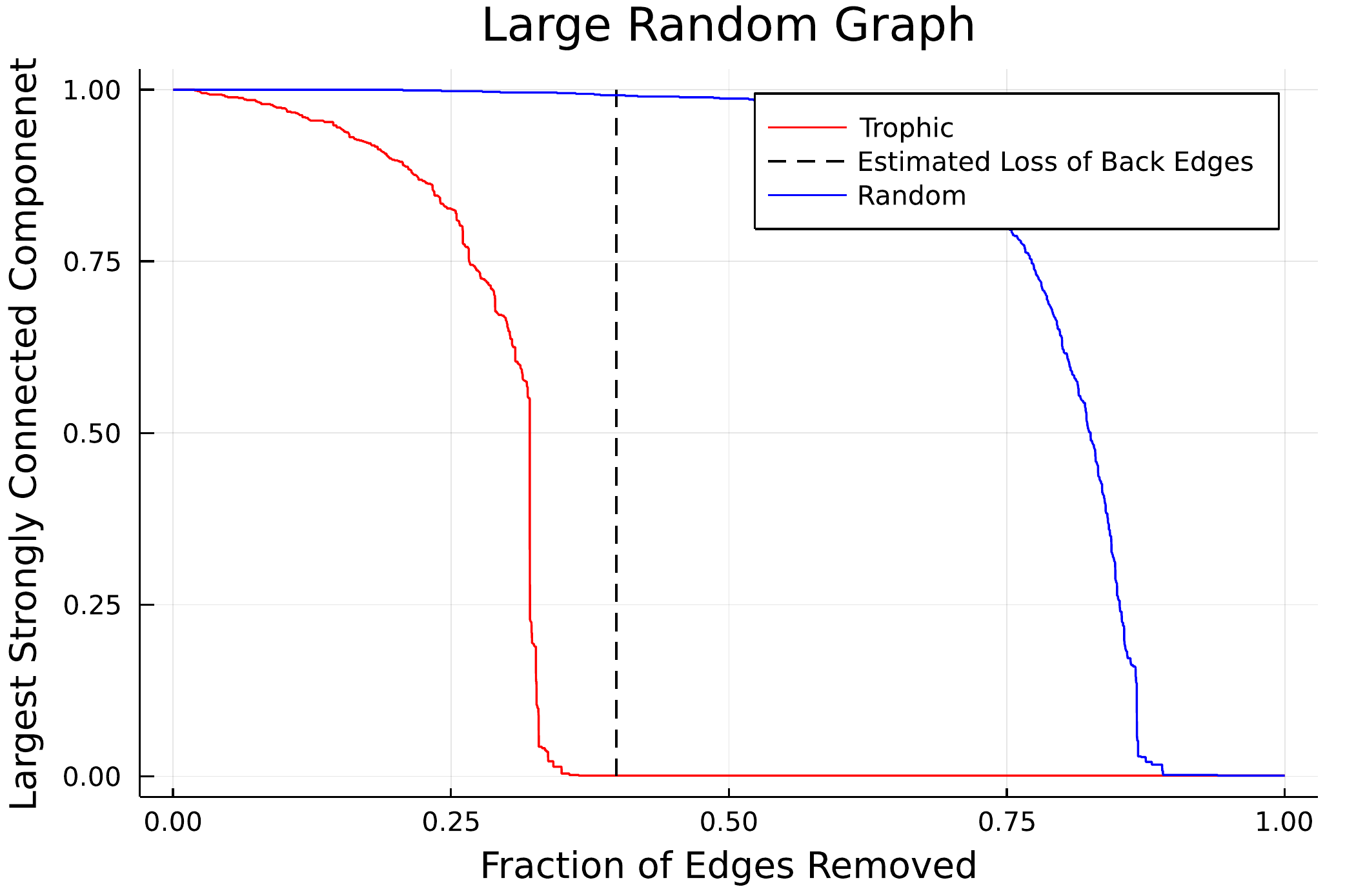}
        	\caption{  Erd\H{o}s–R\'{e}nyi graph. $N=1000$, $\langle k \rangle =10$.  
        	}
        	
        \label{fig:Target_ER}
        \end{subfigure} 
        \caption{ Size of the strongly connect component as edges are removed randomly and in order of Trophic Difference.}
        
\end{figure}

To demonstrate how the strong connectivity of a network depends on the edges which break the hierarchy we can conduct a targeted attack on those edges and compare the degradation of the strongly connected component to that observed in a random attack. We remove edges in order of their trophic difference, starting by removing the edges with the most negative trophic difference. This only takes into account the hierarchical organisation, it may be possible to destroy the strongly connected component faster using a different method, for example attacking bottleneck edges or specifically trying to target edges breaking the hierarchy in different components of the network. However when all the backwards edges are removed all cycles are destroyed and the strongly connected component is guaranteed to vanish. This can be demonstrated in real networks, figure  \ref{fig:Target_C_elegans}, such as the connectome of the worm, \textit{C.Elegans}, which is the only organism to have this fully mapped. The point at which we estimate the ``backwards'' edges to vanish, shown by the dashed line in figure \ref{fig:Target_C_elegans}, is analytically estimated from equation \ref{back_edge_eq} and predicts well the point the strongly connected component vanishes completely.

Similar results can be found for networks where there is no expectation of this kind of organised structure like a dense random graph. This is shown in figure \ref{fig:Target_ER}. The figure shows that even in networks where there is little structure expected there still exists some sort of directional organisation that can be exploited to break down the network. The number of edges needed to make the network acyclic and have zero strongly connected component depends on the total number of backwards edges which is a function of the Trophic Incoherence and the degree as in the percolation transition above. This demonstrates that the strongly connected component can be understood by focusing on the backwards edges as was motivated in the above derivation.

One important thing to note about this targeted attack is that it does not generally affect the size of the weekly connected component until after all the backwards edges are removed and then it behaves in the same way as a random attack in most real-world cases, for more detail see Appendix \ref{Weak_connectivty}. This could be useful in situations where you want to attack the strongly connected component without disrupting the weakly connected component which would happen if bottleneck edges were attacked for example. This means that the backwards edges can act as an approximation for the feedback arc set \cite{Sun2017BreakingHierarchies}. Trophic level may not perform as well as specialist methods at this task \cite{Sun2017BreakingHierarchies} but does provide an analytical estimate of how many edges you expect to need to remove in any large network which can be simply calculated from F. It explains why certain networks are more difficult to render acyclic based on where they lie on the phase diagram.  

\subsection{Importance of Attacks on Backwards Edges for Dynamical Processes} \label{dyanmics}

\begin{figure}[H]
     \centering
      \begin{subfigure}[b]{0.48\textwidth}
         \centering
         \includegraphics[width=\textwidth]{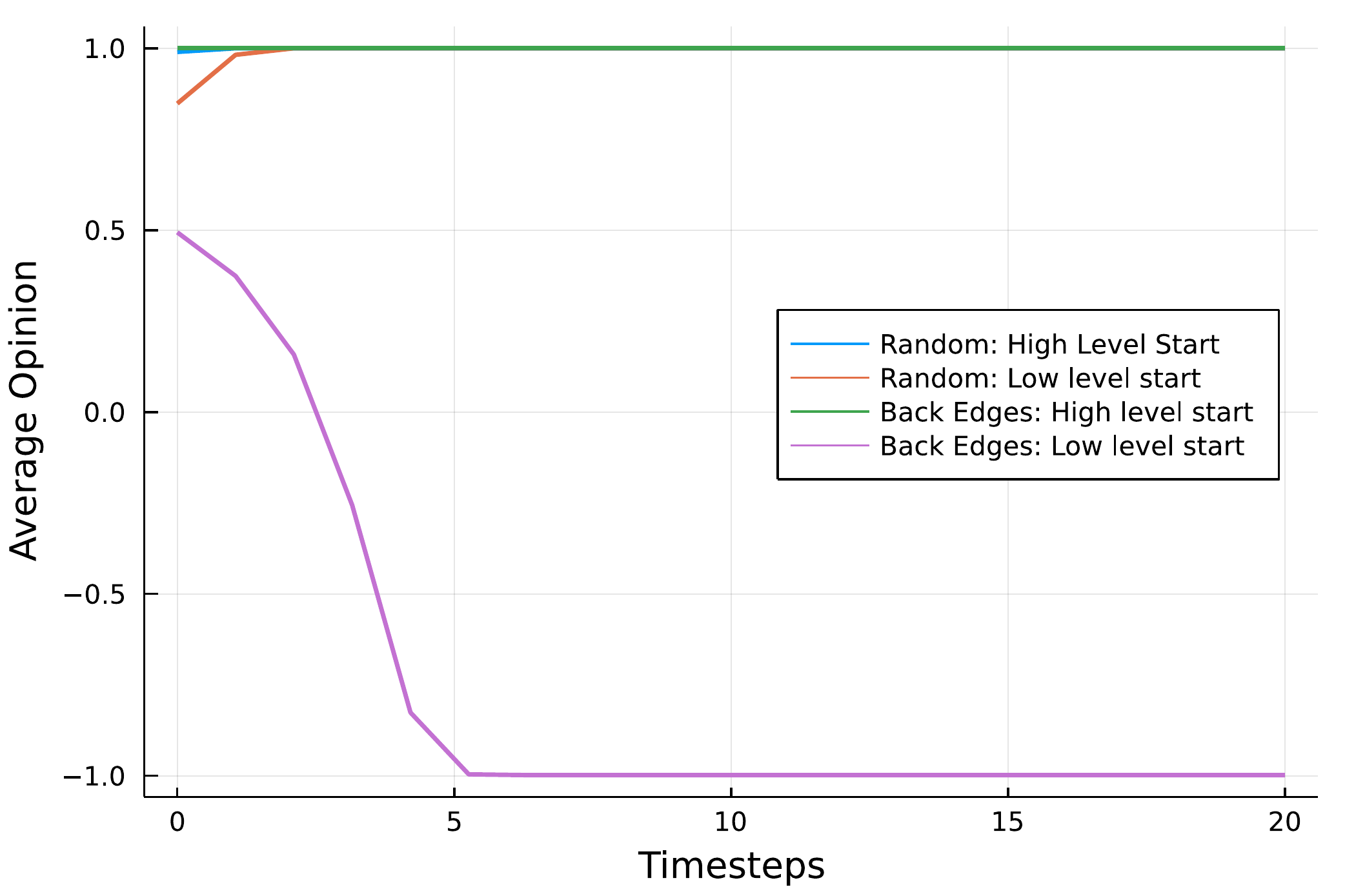}
         \caption{Majority Vote Dynamics with Opinion Inertia on ER random graphs with $N=1000$ and $\langle k  \rangle = 10$ with 20\% of most backwards edges targeted or random edges targeted with different starting locations in the hierarchy of with a fifth of nodes taking a new opinion.}
         \label{fig:Maj_Vote}
     \end{subfigure}
     \hfill
     \begin{subfigure}[b]{0.48\textwidth}
         \centering
         \includegraphics[width=\textwidth]{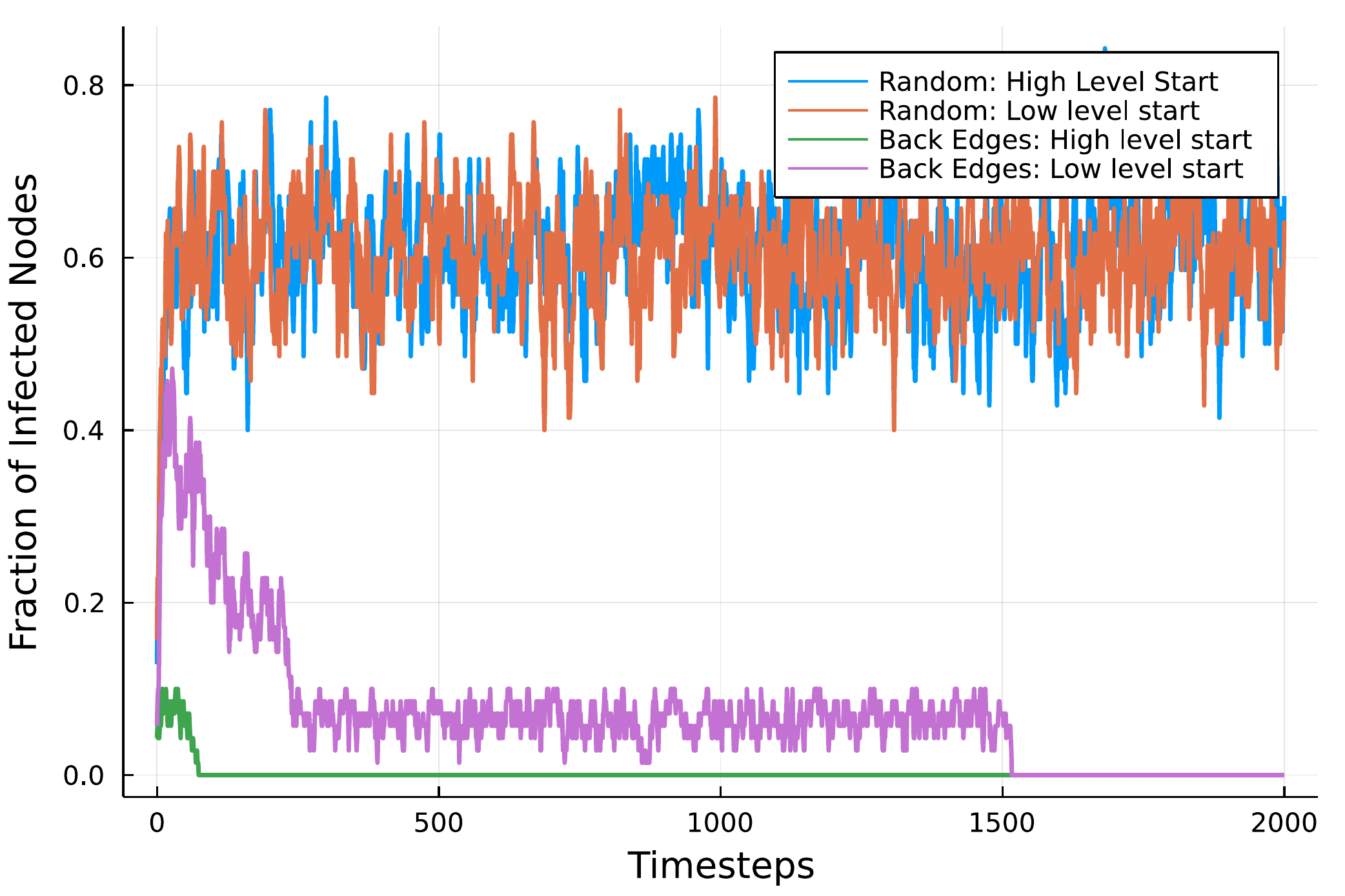}
         \caption{Spread of infection with time in SIS model on High School Social Network \cite{Rossi2015TheVisualization} with 20\% of most backwards or random edges targeted with different starting locations in the hierarchy and an initial infection in 5\% of the students. $pI = 0.2$, $pR= 0.1$.}
         \label{fig:SIS}
     \end{subfigure}
     \hfill
     \begin{subfigure}[b]{0.48\textwidth}
         \centering
         \includegraphics[width=\textwidth]{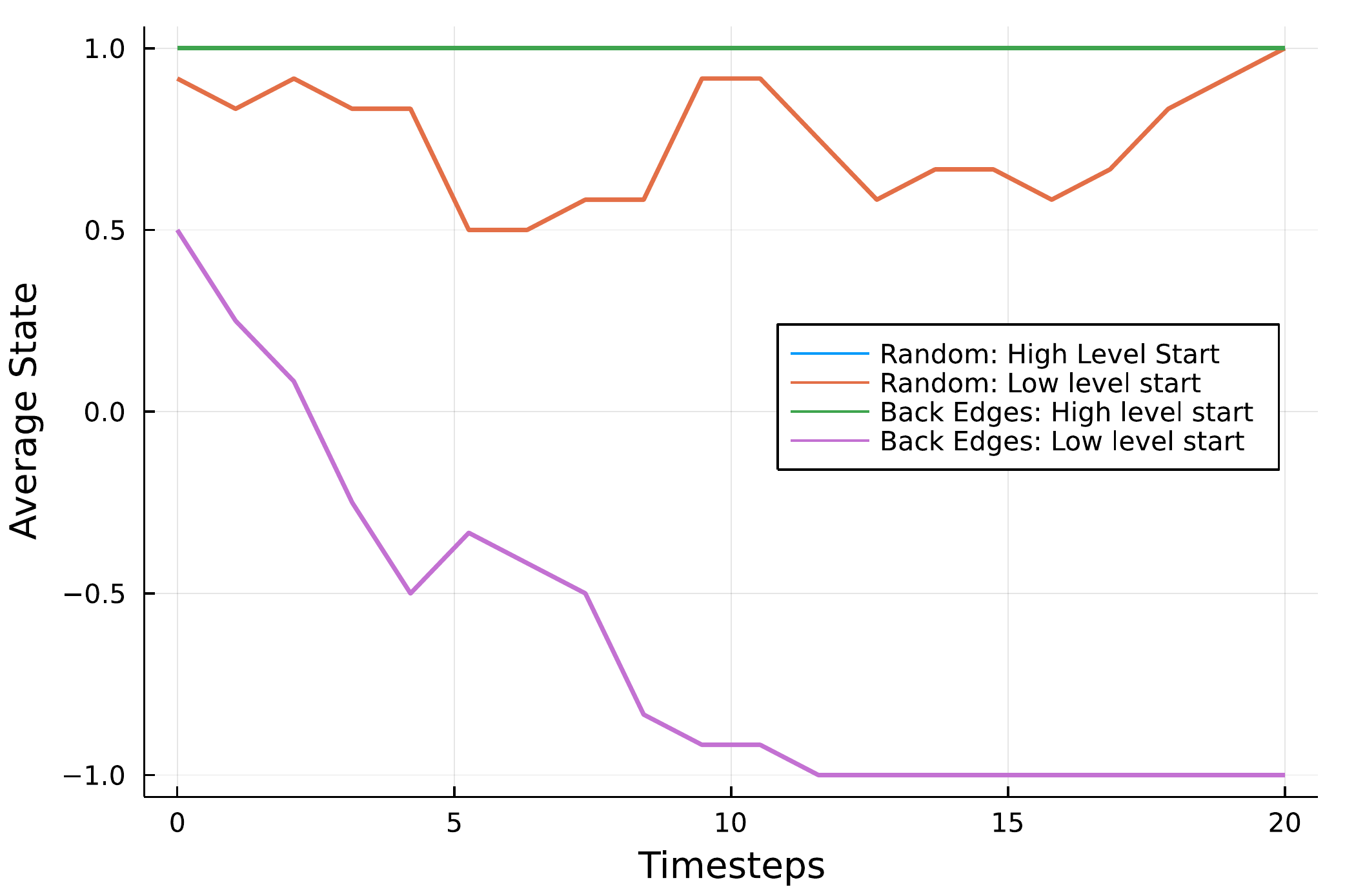}
         \caption{Voter Model on Trade Network \cite{DeNooy2018ExploratoryPajek} with 33\% of most backwards or random edges targeted  with different starting locations in the hierarchy and 5\% of nodes taking the new state. }
         \label{fig:Voter_model}
     \end{subfigure}
     \hfill
     \begin{subfigure}[b]{0.48\textwidth}
         \centering
         \includegraphics[width=\textwidth]{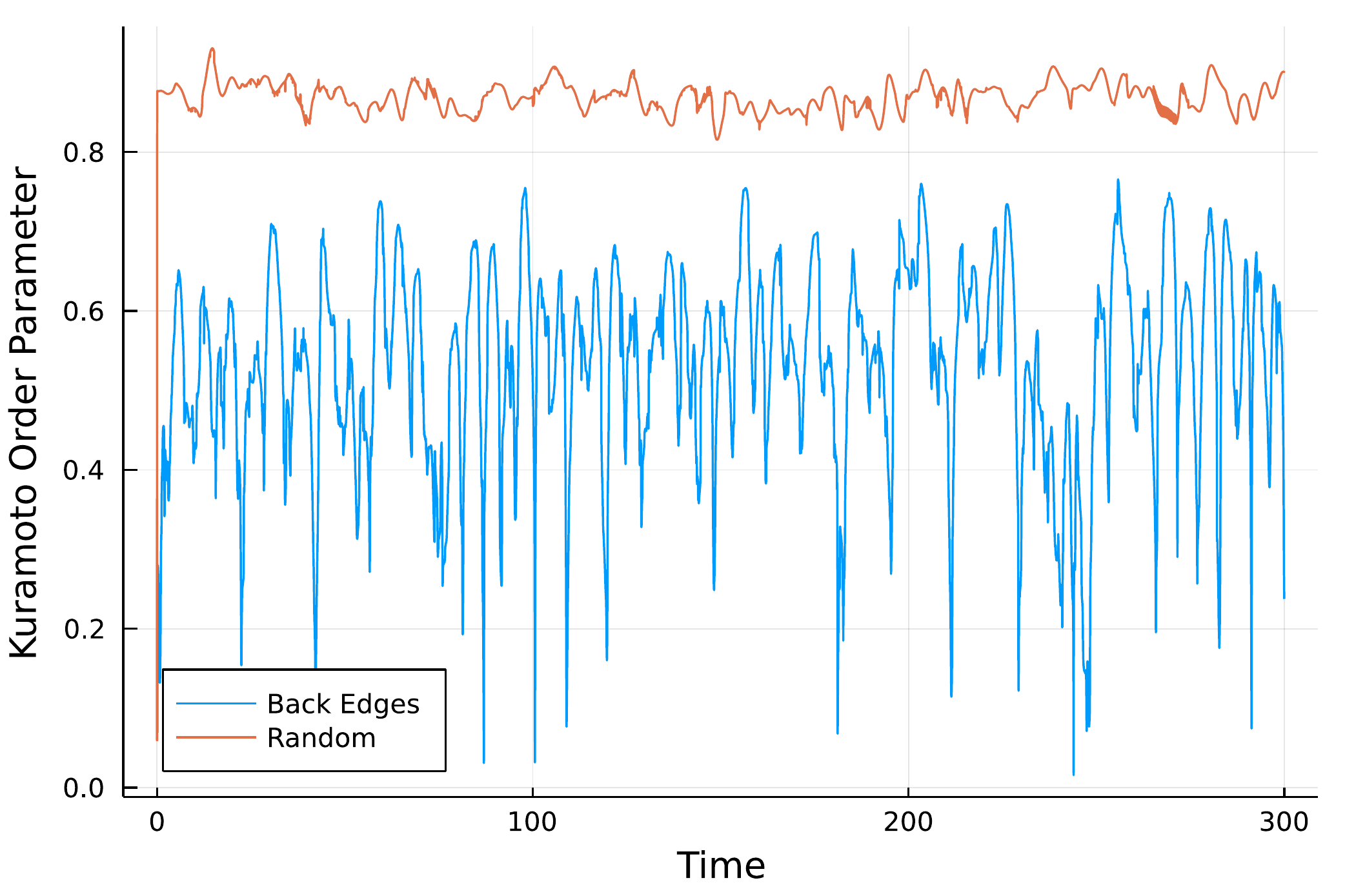}
         \caption{Synchronisation of Continuous Kuramoto Oscillators on the \textit{C.Elegans} connectome \cite{White1986Elegans} with 20\% of most backwards edges targeted and random edges targeted starting from random phase between zero and $2\pi$. }
         \label{fig:Kuramoto}
     \end{subfigure}
        \caption{Dynamics of Real-World Networks after targeted attack on Backwards edges and Random Edge Attacks. All real-world data sets can be found at \cite{DataSamJohnson} or \cite{DeNooy2018ExploratoryPajek} for the trade network.}
        \label{fig:Dyanmics Exampls }
\end{figure}

Strong connectivity is very important for many real-world networks and their dynamics. For illustrative purposes we demonstrate how a spreading process defined by a Susceptible-Infected-Susceptible (SIS) model dynamics; Opinion formation governed by Majority Vote Dynamics; changes in states of nodes governed by the the Voter Model and Synchronisation of Continuous Kuramoto Oscillators all change after a targeted attack on the backwards edges calculated using Trophic Analysis and random attack. These are all demonstrated in figure \ref{fig:Dyanmics Exampls }.  

Majority vote is a very simple model of opinion formation which is demonstrated in figure \ref{fig:Maj_Vote}. Each agent in the model is given an opinion and then updates their opinion if the majority of their neighbours have a different opinion. It is very simple but shares some similarities with other discrete  dynamics widely used in complex networks such as the SIS epidemic model \cite{Volkening2020ForecastingInfection}, Hopfield neural network model \cite{Rodgers2022NetworkNetworks}, the Ising model of magnetisation \cite{Kim2021Majority-voteNetworks} and the Moran process describing evolution of populations \cite{Moinet2018GeneralizedAttractiveness}.  The simplest way to define majority vote dynamics is a system where there are two opinions denoted as +1 and -1. Each agent holds either of the two options and updates their state according to the majority opinion of their neighbours. In the case of a draw the new state can be randomly selected or remain in the previous state. We choose to fix the system in the previous state to prevent random opinions filtering through the system from the nodes with in-degree 0 when the backwards edges are removed. This update rule can be written as  \begin{equation}
 S_{i}(t + \Delta t) = \text{sgn}\left(\sum_j A_{ij}S_j(t)  + \delta S_i(t) \right).\end{equation}
Where $\delta$ is a small positive constant less than one to account for balanced opinions. We also update the state of the system in parallel. In this setup, for maximum simplicity, we take an Erd\H{o}s–R\'{e}nyi random graph, with $N=1000$ and $\langle k \rangle =10 $, and give one fifth of the nodes a new opinion, labelled -1, and observe how that opinion spreads over time. This could represent a political belief or taking part in a social trend. Figure \ref{fig:Maj_Vote} shows that how the opinion forms depends on where in the hierarchy the opinion starts and how much feedback there is in the system driven by the backwards edges. When the new opinion starts at the bottom of the hierarchy and the backwards edges have been attacked it can spread through the whole system and it becomes a majority opinion amongst the nodes. When the opinions starts at the bottom of the hierarchy after a random attack it can still slightly spread up through the system but since there is still feedback in the system the affect of the new opinion is damped so it quickly dies out. When the opinions starts at the top of the system and the back edges have been attacked it can not propagate back down the system so the new trend quickly dies out. When the opinions starts at the top of the system after a random attack it is also quickly replaced by the opinion from the lower trophic level nodes. This also demonstrates that hierarchical structure can be found and exploited even in networks like random graphs where little structure is expected.

The SIS model, figure \ref{fig:SIS}, is  a simple spreading process which could be imagined to represent in the simplest way the spreading of an infection which you can catch multiple times and lack immunity like the common cold or sexually transmitted disease; the spreading of a meme in a social network; selling in a trading network or activation in a neural network. In this model each node can either be susceptible or infected. If a node is susceptible it becomes infected if any of its in-neighbours are infected with probability $pI$, at each time step. If a node is infected it loses the infection and transitions to the susceptible state with probability $pR$. There are many possible variations of this model however we use the simplest case for demonstration \cite{Hethcote1989ThreeModels}. Using parallel updates we start the system with 5\% of the nodes infected and the rest susceptible. This is shown in figure \ref{fig:SIS} where the dynamics take place on a High School Social Network. If random edges are removed the backwards edges remain and then the infection can cycle round the network and the infection becomes endemic. If the network is strongly connected the final state is not affected by where in the the network the infection appears however the initial spread can be affected by where in the hierarchy it starts. When the backwards edges are removed there is no strongly connected component to maintain the infection so it eventually dies out. However the hierarchy induces an asymmetry in the network. If the infection starts in the low trophic level nodes it can spread through a large part of the network before it dies out so many nodes see the infection while if the infection begins in the high level nodes it has nowhere to spread to so quickly dies out, figure \ref{fig:SIS}. This is very important to understand as each of these scenarios can have very different consequences depending on the nature of the spreading agent and the system in question.  

A voter model, figure \ref{fig:Voter_model}, is another simple model which can represent opinion formation or general updating of states of the agents in a network \cite{Redner2019Reality-inspiredMini-review}. We take two discrete states labelled $+1$ and $-1$ and update the system in parallel such that at each time step a node selects one of it's in-neighbours at random and copies the state of the chosen neighbour. This model is very simple however variants of this model can be used in modelling real voting processes \cite{Fernandez-Gracia2014IsVoters}, economics \cite{Kirman1993AntsRecruitment} and chemistry \cite{Fichthorn1989Noise-inducedModel,Considine1989CommentModel}. In our example dynamics we simulate the simple voter model on a trade network. Where the new state can represent any relevant binary change in the function of an entity for example if the agent suffers a delay or shortage in production or begins selling off particular assets. We start the system with 5\% of the nodes in the new state and destroy one third of the edges, more edges have to be attacked due to the network being small and dense as well as quite incoherent. When the perturbation is made to the high level nodes the new state can not take hold and the system maintains it's previous state, figure \ref{fig:Voter_model}. However when the new state is introduced to the low level nodes it can gain a foothold. When it is presented at the low level nodes after a random attack it survives for some time before disappearing while after an attack on the backwards edges the new state is able to overtake the entire system.

The Kuramoto model is a very important model of synchronisation \cite{Acebron2005ThePhenomena} used in a wide variety of setting in particular in Neuroscience \cite{Bick2020UnderstandingReview, Cumin2007GeneralisingBrain}. In figure \ref{fig:Kuramoto} we simulate Kuramoto oscillators on the neural structure of the nematode \textit{C.Elegans} after random and targeted attack. It can be treated analytically in simple network topologies but in complex networks the model must be solved numerically. We use NetworkDyanmics.jl \cite{Lindner2021NetworkDynamics.jlComposingJulia} to solve the system of differential equations used in our variant of the model. Each oscillator has a phase, $\theta$, which evolves according to the equation \begin{equation}
    \frac{d\theta_i}{dt } = \frac{K}{k_{i}^{\text{in}}}\sum_{j=1}^N A_{ji}\sin{(\theta_j - \theta_i)} + \omega_i.
\end{equation}   
Where $K$ is the coupling constant, $k_{i}^{\text{in}}$ is the in-degree of node $i$ and $\omega_i$ is the natural frequency of node $i$. We use the form normalised by in-degree so that the oscillators update at similar rates even if they have many input nodes. $K$ is taken to be 50 to ensure synchronisation and the natural frequency of each node is drawn from a normal distribution with mean 0 and standard deviation 1. The synchronisation of the oscillators is shown in figure \ref{fig:Kuramoto} and measured by the order parameter \begin{equation}
    r = \frac{1}{N} \lvert \sum_{i=1}^N  e^{i\theta_i} \rvert.
\end{equation} 
This reaches 1 when all the oscillators are fully in phase. Starting from each oscillator having an initial phase between 0 and $2\pi$ under a random attack the system is still able to synchronise. However when the backwards edges are attacked the synchronisation is less strong, \ref{fig:Kuramoto}, and the system is disrupted.

These results tie into existing literature on directionality, spreading \cite{Zhu2014InfluenceNetworks} and epidemic thresholds in directed networks \cite{Li2013EpidemicNetworks}. Where the epidemic threshold can be considered a function of the spectral radius and driven by the directionality as measured by the fraction of bidirectional edges \cite{Li2013EpidemicNetworks}. Our results are in agreement with this as a bidirectional edge pair must count at least one edge where the trophic level difference is less than or equal to zero. We extend the definition of directionality to make it more general and complete rather than simply the fraction of edges which are bidirectional \cite{Li2013EpidemicNetworks}. Our results can also be restated in terms of the spectral radius of the adjacency matrix as there is an analytical estimate of the spectral radius as a function of $F$ \cite{MacKay2020HowNetwork}. Similar results demonstrating the affect of directionality and hierarchy on the performance of Hopfield-like neural networks can be found in \cite{Rodgers2022NetworkNetworks}. The results demonstrate the importance of the backwards edges to the system across a variety of dynamics and scales which we expect to hold in a variety of other systems. However it should be stressed that the importance of the backwards edges to dynamics depends on a variety of factors. Firstly the impact of attacking the strongly connected component depends on the initial size and distribution of that component as if the component is initially very small destroying it may have little affect. Degree distribution can also interplay very strongly with dynamics both in the sense that hub nodes of very high degree can play in important role in controlling the dynamics and that directed networks can potentially have many nodes which have in-degree zero \cite{Wright2019TheDynamics}. Nodes which have in-degree zero have no input from the system so internally set their own state so can play a large role in controlling the system dependant on their placement in the hierarchy, out-degree and the specifics of their internal dynamics. In addition this could be repeated with another measure of the hierarchical ordering and targeting of the backwards edges however you could not analytically estimate the number of edges you would need to target before enumerating all the backwards edges and would also lack the link to the global directionality.

\section{Discussion and Potential Applications}

There are a wide range of network applications where percolation in networks has been observed to be important \cite{Li2021PercolationApplication}. We highlight a few areas where our work may be useful but this is not necessarily exhaustive. Strong connectivity and percolation can play an important role in city planning as networks of one-way streets must be strongly connected \cite{Verbavatz2021FromGraphs}. Using Trophic Analysis to study network connectivity may be useful in ecological settings \cite{Schick2007DirectedNetwork} with a particular focus on spreading process. Trophic Analysis is suited towards spreading processes where the network is directed and there is some ordering to the network structure. A real-world example of this is the spreading of crown-of-thorns starfish on coral reefs \cite{Hock2014ConnectivityReef,Hock2016ControllingPopulations,Hock2017ConnectivityReef}. These starfish are a pest which eat coral reefs and can damage ecosystems. Outbreaks are governed by the spread of their larvae by the ocean currents. This process is directed as the larvae move in the direction the ocean flows. This is the kind of process Trophic Analysis could lend itself to as it could be used to understand the global connectivity structure to see if the outbreak is likely to spread across the reef or in a directed manner. It can be used to extend the existing analysis of a regions vulnerability to outbreaks or danger as a  starting point beyond simply the size of the out and in components \cite{Hock2014ConnectivityReef} but factor in where reefs sit in the network hierarchy determined by trophic level. Our work could also relate to the growth of biological neural networks and formation of a giant strongly connected component of cells \cite{Breskin2006PercolationNetworks,Soriano2008DevelopmentCultures}. These neurons have previously been grown in circumstances where there is limited hierarchical structure and the degree distributions are well know however if the cells were exposed to a directed gradient or in a real life system and more likely to grow in a particular direction Trophic Analysis may play a role in explaining this percolation threshold. Our results may also partially explain the difference in percolation thresholds for dynamical processes on directed networks \cite{Wright2019TheDynamics} compared to the undirected case due to the effect of hierarchical ordering increasing the threshold for the strong connectivity. This will be observed even in random graphs as some hierarchy can still be observed and trophic incoherence does not reach one. The percolation of the strongly connected component and the direction of flow and spread of information may also play a role in communication networks and control and decision making in organisations \cite{Pilgrim2020OrganisationalAnarchy}. In general Trophic Analysis can be used to modify the dynamics by understanding the hierarchical organisation and the effect of localised perturbations as well highlighting the role of hierarchy breaking edges in driving strong connectivity, feedback and resilience in complex systems. Trophic Analysis is also a useful metric due to the simplicity of the calculation and the interpretability as the notion of directionality and place in hierarchy is quite simple and intuitive. This makes the method attractive to be employed in many settings as the barrier to entry is relatively low while still providing good intuition into the structure of a directed network. Additionally it provides motivation to focus on the intrinsic directional aspects of real-world systems \cite{Johnson2020DigraphsSystems} which are understudied compared to the undirected case. Our results for the strong connectivity despite the success are also a very simple approximation of the percolation of the backwards edges so it may be possible to extend or repeat this result with a different measure of hierarchy or more information about the degree distribution.

\section{Conclusion}
In conclusion, we have shown how by using Trophic Analysis to study the hierarchical ordering and global directionality it is possible to analytically estimate the number of ``backwards'' edges which break this ordering and estimate the threshold for the network to be strongly connected by using percolation theory on these edges. From this a phase diagram of strong connectivity in terms of Trophic Incoherence and mean degree can be derived which holds well for real directed networks. This shows that strong connectivity in directed networks is driven by more than just the degree and the hierarchy can play a significant role. We highlight these results by conducting a targeted attack on ``backwards'' edges and by implementing a range of dynamics where the behaviour is dictated by the strongly connected component and start location in the hierarchy.

\section*{Acknowledgements}

All graph manipulations were carried out using Julia Package Graphs.jl \cite{Graphs2021}. The authors would like to thank the Centre for Doctoral Training in Topological Design and Engineering and Physical Sciences Research Council (EPSRC) for funding this research. SJ also acknowledges support from the Alan Turing Institute under EPSRC Grant EP/N510129/1. Author Contributions: NR wrote the initial manuscript, came up with the initial concept, did the calculations and derivations. PT and SJ supervised, advised the process, suggested ideas and edited the manuscript.  

\appendix

\section{Removal of Backwards Edges and Weak Connectivity} \label{Weak_connectivty}

\begin{figure}[H]
      		\centering
            \includegraphics[width=0.5\linewidth]{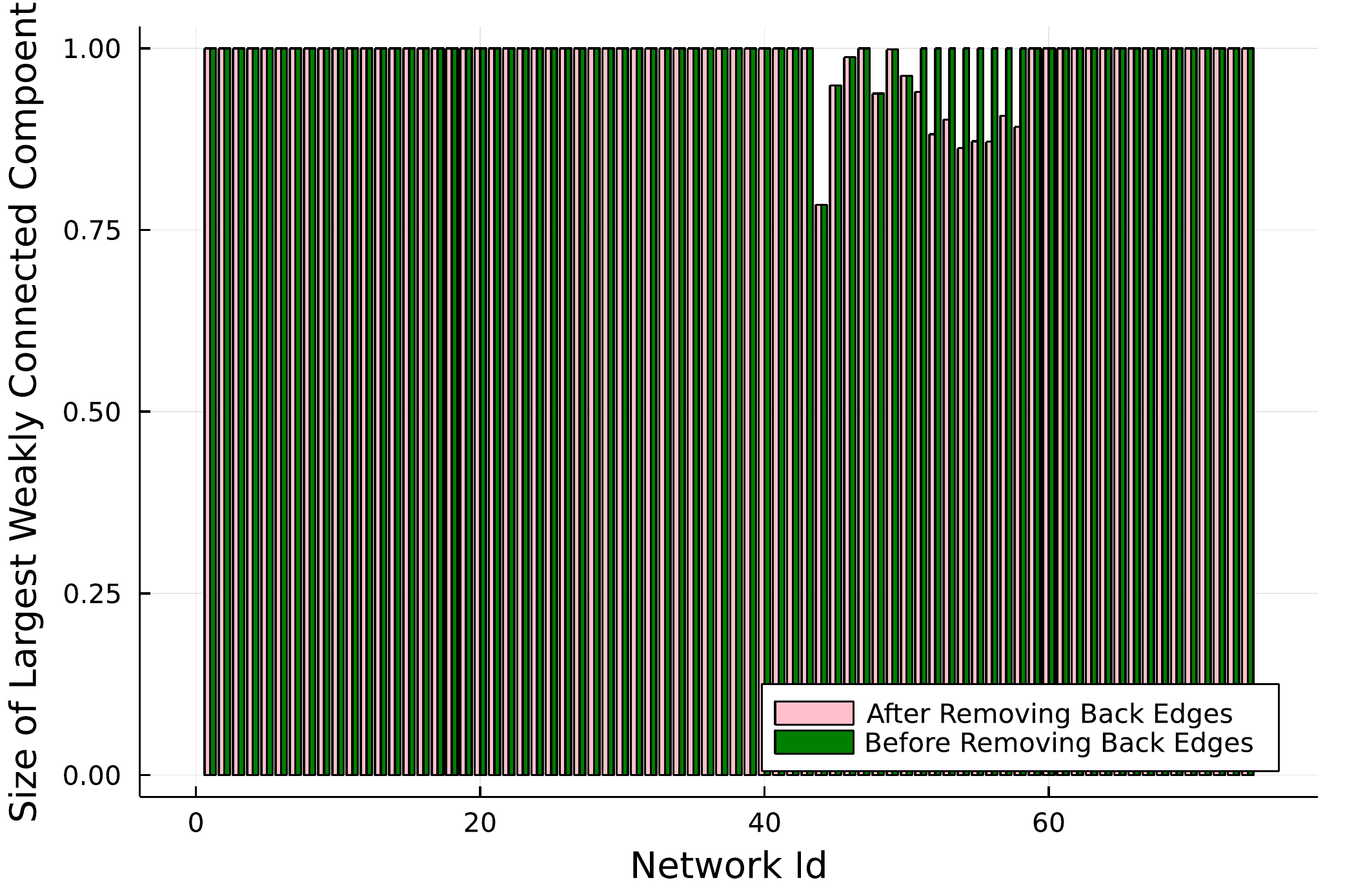}
        	\caption{ Lack of change in the size of the weakly connected component as the backwards edges are removed from real networks \cite{DataSamJohnson} 
        	}
        	
        \label{fig:Weak_connectivty_all}
        \end{figure}

Backwards edges are in general linked to cycles and reciprocal edges this means than in general removing one backwards edge is unlikely to separate a network into distinct components. This explains why the size of the weakly connected component is generally unaffected when the backwards edges are removed in real networks, figure \ref{fig:Weak_connectivty_all}.

\begin{figure}[H]
      		\centering
            \includegraphics[width=0.8\linewidth]{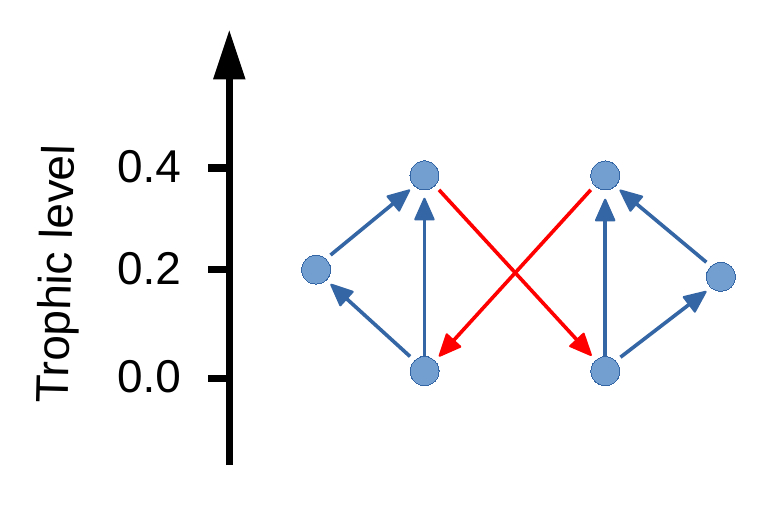}
        	\caption{ Example of network which is connected by two edges which go backwards in Trophic level.   
        	}
        	
        \label{fig:Counter_example}
        \end{figure}

However, there are specific structures composed of interlocked cycles which can cause the network to become disconnected if the trophic level is calculated only once and then all the edges which are initially backwards are removed as shown by the example in figure \ref{fig:Counter_example} where the backwards edges are highlighted in red. This however is a specific case and does not seem to be found when studying the connectivity of real networks. In addition if the trophic levels were recalculated after the first edge was removed the second edge would then become forwards. This leads to the property that if the trophic level is recalculated then the network will not become disconnected. This is proved below.

Given an initial graph G(V,E) at each step remove the edge which is most backwards in trophic level and then recalculate the trophic level. Stopping when all the edge differences are non-negative. In order to break the network into separate components via this method there would require a situation where the graph has been separated into two disjoint components joined by a single edge which upon removal would break the graph into disconnected pieces. If the trophic levels are recalculated then it is never optimal for this edge to go backwards in trophic level as the levels of one of the components can be modified by a constant which will not change the global coherence apart from across the joining edge where it will become positive. If the edge is backwards it is always possible to reduce incoherence in this way so the configuration can not exist upon recalculation.

%comment out to get back unsrt style and use bibtex
%\printbibliography %Prints bibliography

\renewcommand\bibnumfmt[1]{#1.}

\bibliographystyle{ScienceAdvances}
\bibliography{references,ref1}
%\printbibliography

\pagebreak

\begin{center}
\textbf{\large Supplemental Materials: Strong Connectivity in Real Directed Networks}
\end{center}
%%%%%%%%%% Merge with supplemental materials %%%%%%%%%%
%%%%%%%%%% Prefix a "S" to all equations, figures, tables and reset the counter %%%%%%%%%%
\setcounter{equation}{0}
\setcounter{figure}{0}
\setcounter{table}{0}
\setcounter{page}{1}
\makeatletter
\renewcommand{\theequation}{S\arabic{equation}}
\renewcommand{\thefigure}{S\arabic{figure}}
\renewcommand{\bibnumfmt}[1]{[S#1]}
\renewcommand{\citenumfont}[1]{S#1}
\setcounter{section}{0}

\section{Strong Connectivity By Network Type }

In the main text we give the result for predicting the strong connectivity for real networks where all the networks of different types are shown in the same figure. Here we break down the networks by the categories included in the data set. All the networks are collected in different ways so there may be uncertainty associated with how well networks of each type represent the underlying real world system.

\begin{figure}[H]
      		\centering
            \includegraphics[width=0.8\linewidth]{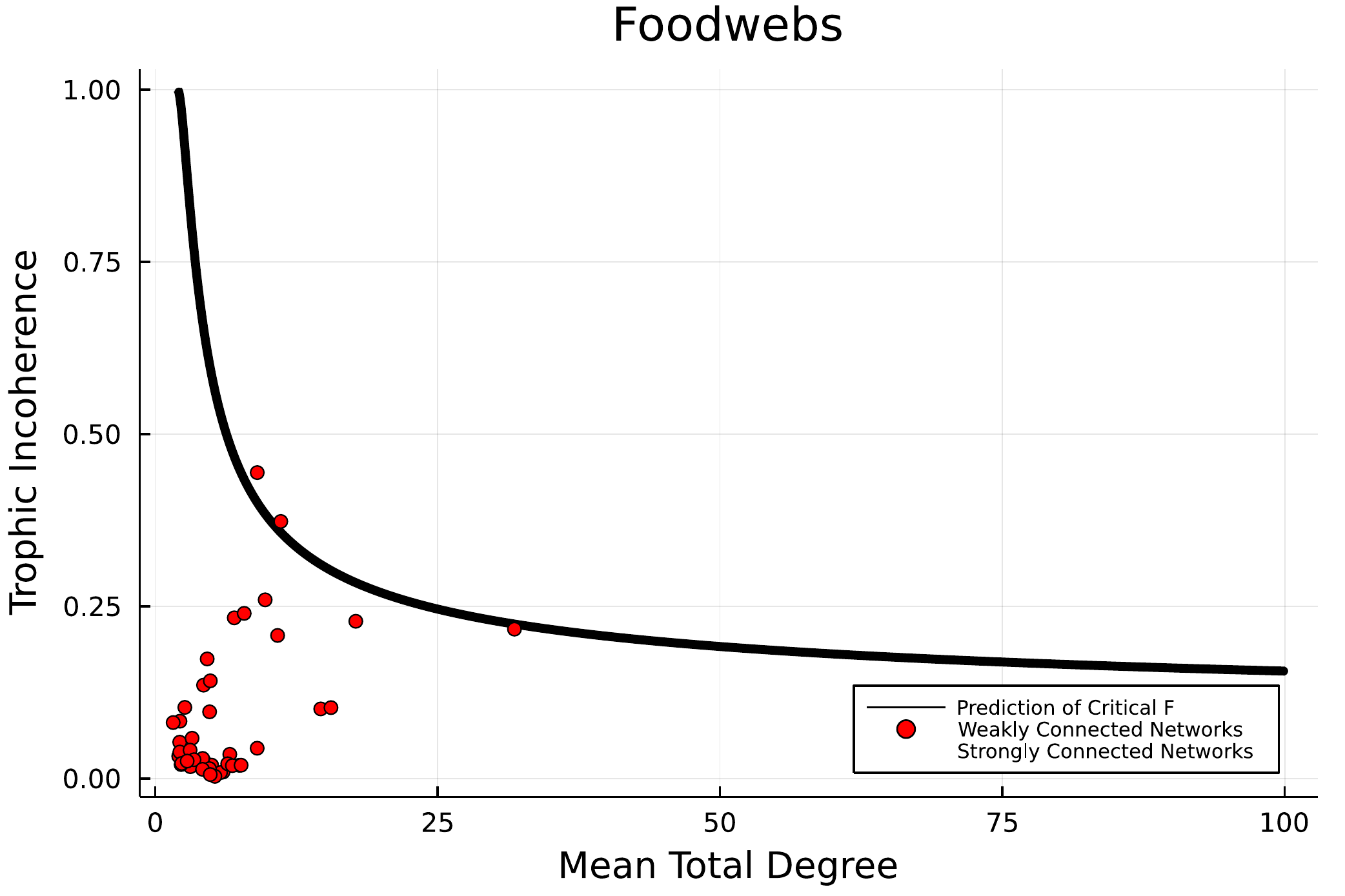}
        	\caption{ Prediction of Strong Connectivity in Food webs from  \cite{DataSamJohnson}}

        \label{fig:strong_foodwebs}
        \end{figure}

Food-webs, figure \ref{fig:strong_foodwebs}, are unlikely to be strongly connected as expected as they have a very hierarchical structure as result of the interactions between species and the flow of energy up the food chain. It is interesting to note that some food webs have high degrees where previous work on random directed networks \cite{Boguna2005GeneralizedNetworks} would expect the network to be strongly connected.    

\begin{figure}[H]
      		\centering
            \includegraphics[width=0.8\linewidth]{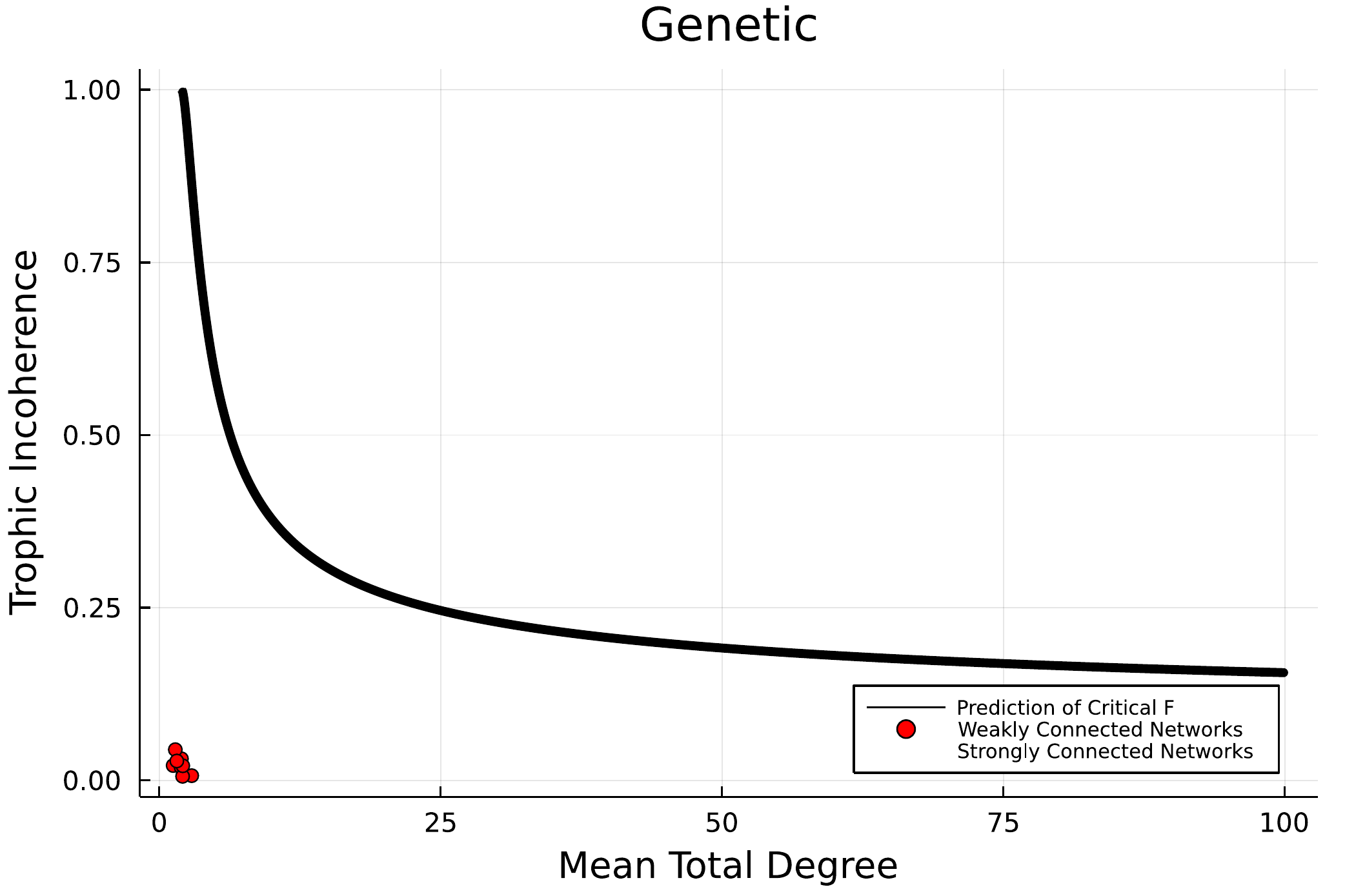}
        	\caption{ Prediction of Strong Connectivity in Genetic Networks from  \cite{DataSamJohnson}}

        \label{fig:strong_genetic}
        \end{figure}

The genetic networks in our data set, figure \ref{fig:strong_genetic}, are all very coherent and lack a strongly connected component. 
 
\begin{figure}[H]
      		\centering
            \includegraphics[width=0.8\linewidth]{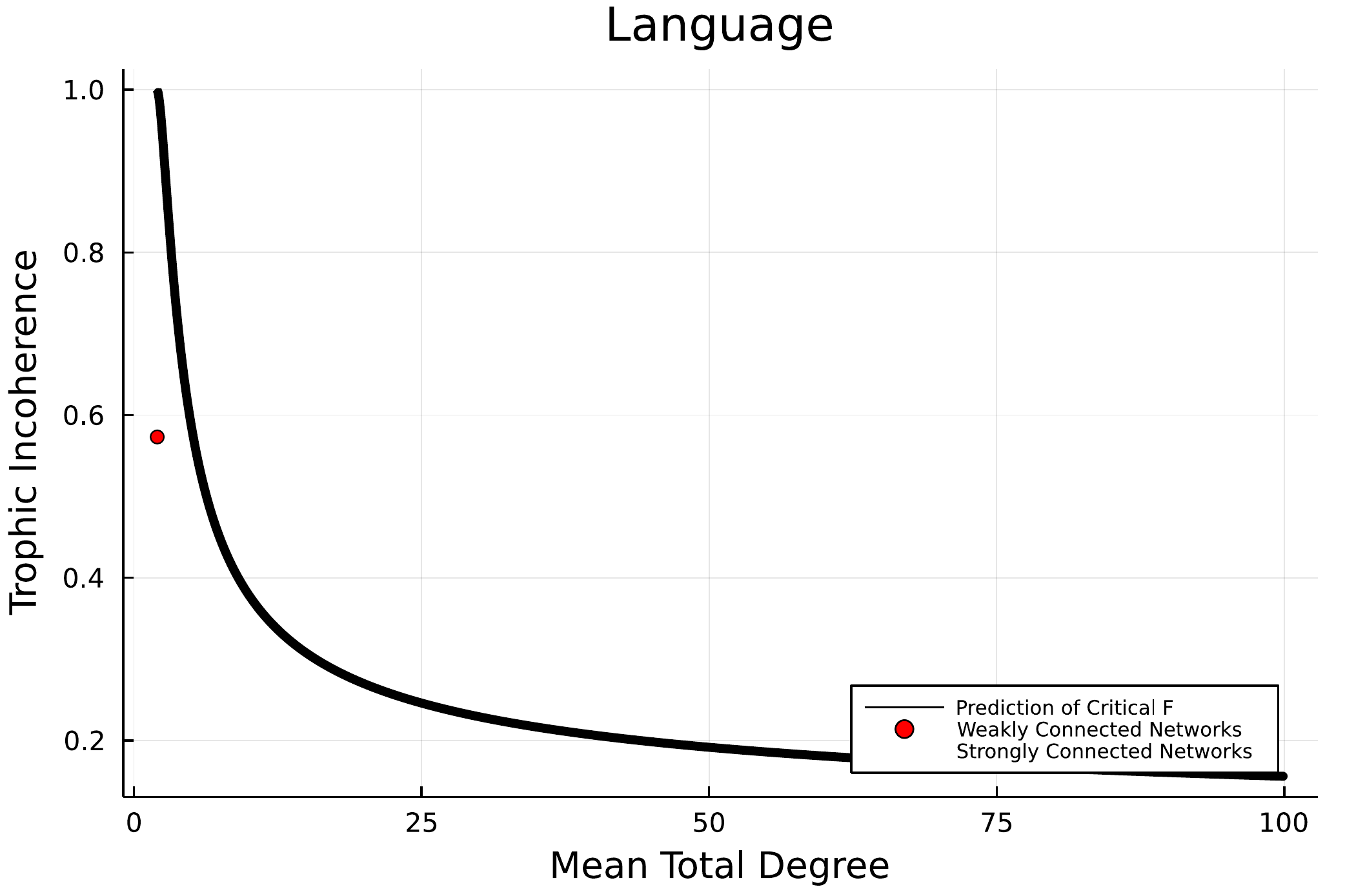}
        	\caption{ Prediction of Strong Connectivity in Language Networks from  \cite{DataSamJohnson}}

        \label{fig:strong_language}
        \end{figure}

 The singular language network, figure \ref{fig:strong_language}, is quite incoherent but lies close to the transition line. 

\begin{figure}[H]
      		\centering
            \includegraphics[width=0.8\linewidth]{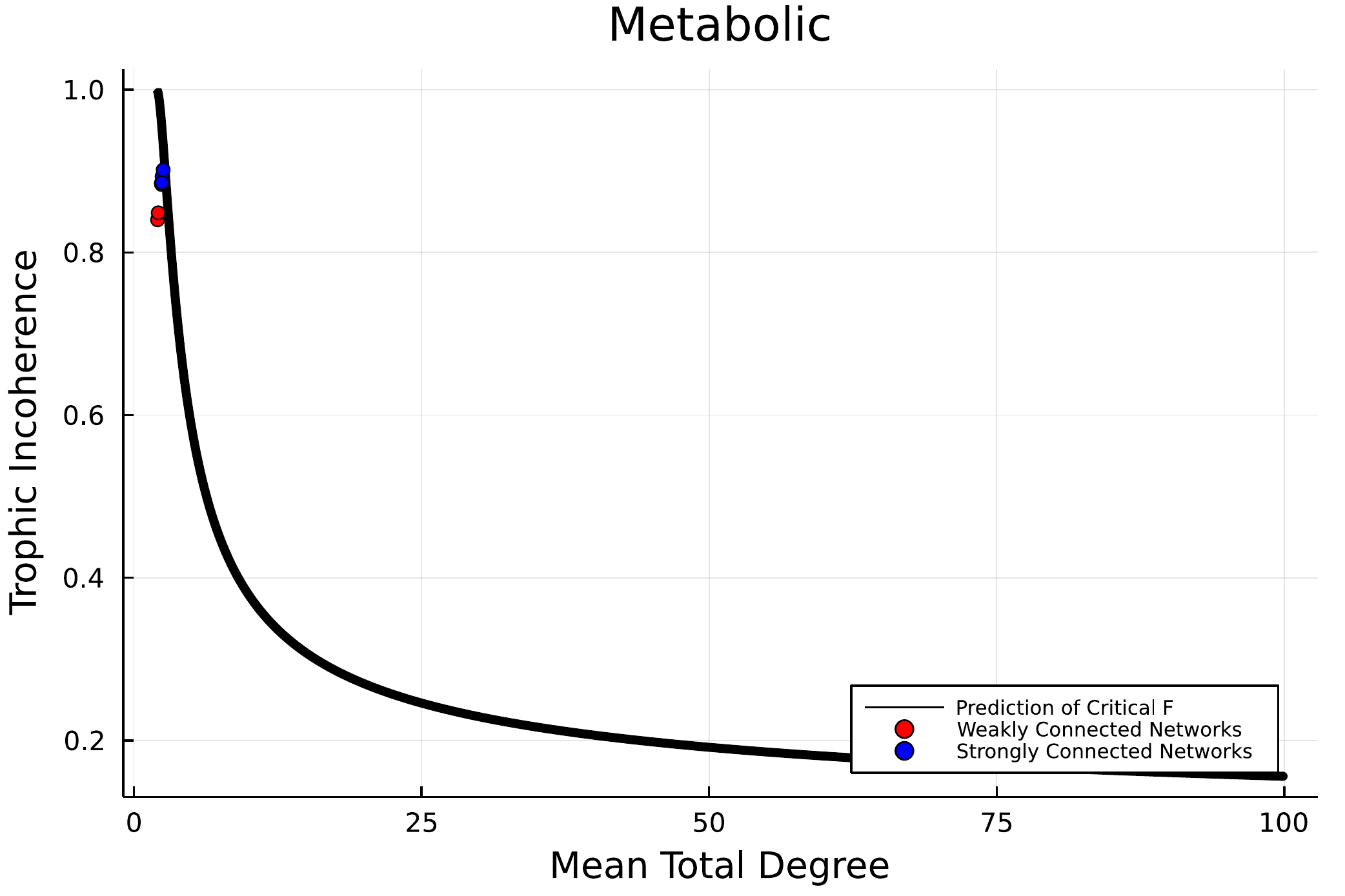}
        	\caption{ Prediction of Strong Connectivity in Metabolic Networks from  \cite{DataSamJohnson}}

        \label{fig:strong_metabolic}
        \end{figure}

 The metabolic networks, \ref{fig:strong_metabolic}, all lie very close to the transition line and are very incoherent so have a little global hierarchical structure but still enough to make our analysis relevant. 
 
\begin{figure}[H]
      		\centering
            \includegraphics[width=0.8\linewidth]{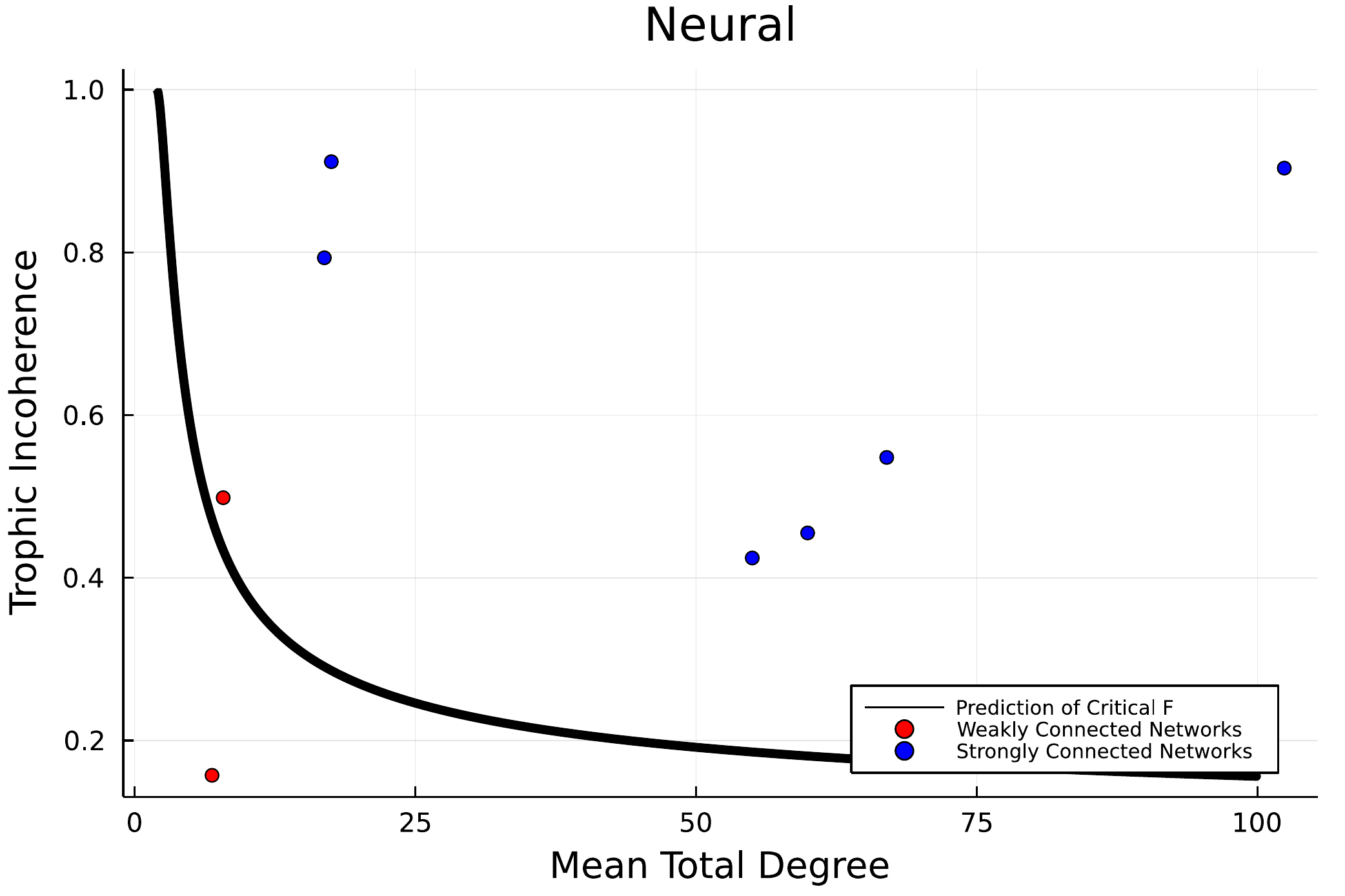}
        	\caption{ Prediction of Strong Connectivity in Neural Networks from  \cite{DataSamJohnson}}

        \label{fig:strong_neural}
        \end{figure}

 There is large variation in the neural network data set, figure \ref{fig:strong_neural}, this is due to the variety of data types and methods employed in this field. This data includes the connectome of \textit{C.Elegans} as well as functional brain networks for a variety of species.
 
\begin{figure}[H]
      		\centering
            \includegraphics[width=0.8\linewidth]{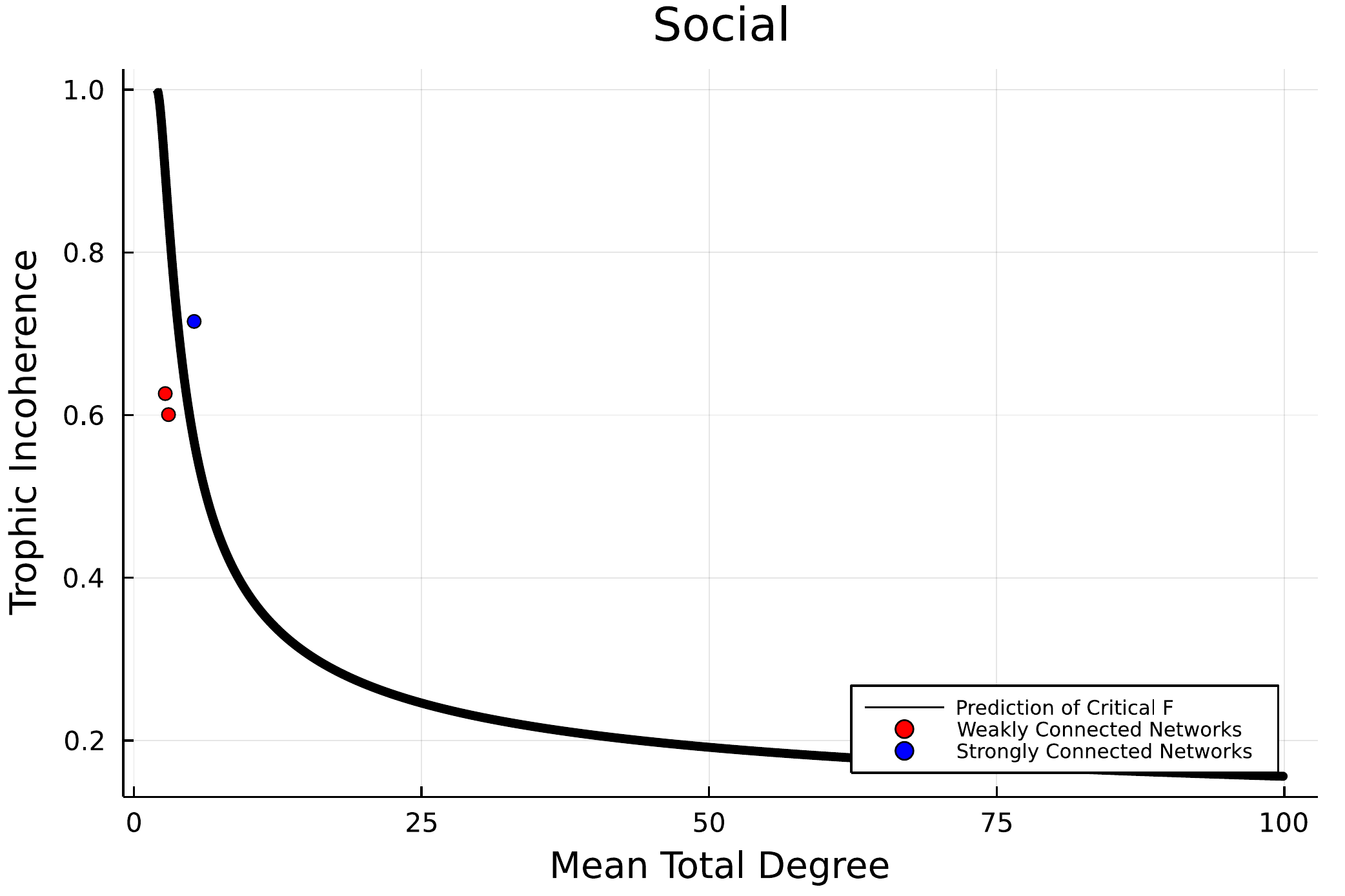}
        	\caption{ Prediction of Strong Connectivity in Social Networks from  \cite{DataSamJohnson}}

        \label{fig:strong_social}
        \end{figure}

 The social network comprised of in-person friendships all lie close to the transition line, figure \ref{fig:strong_social}, between strong and weak connectivity. Again social networks are a type of real network where you might initially expect there to be very little hierarchy and ordering.

\begin{figure}[H]
      		\centering
            \includegraphics[width=0.8\linewidth]{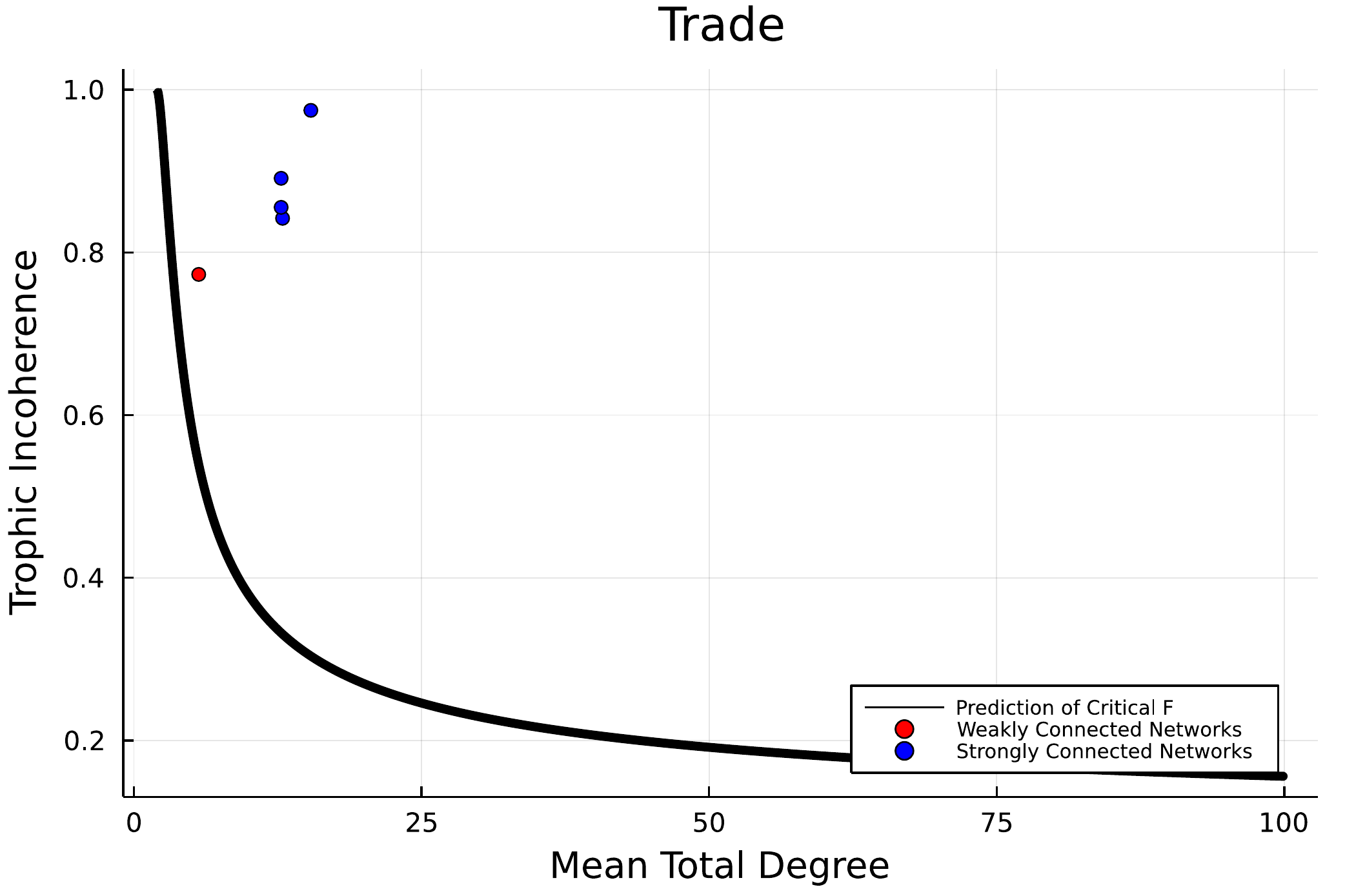}
        	\caption{ Prediction of Strong Connectivity in Trade Networks from  \cite{DataSamJohnson}}

        \label{fig:strong_trade}
        \end{figure}

 The trade networks in our data set, figure \ref{fig:strong_trade}, are all in a similar region which is mostly strongly connected but still exhibits some hierarchy.

\section{Branching Factor for all Networks}

\begin{figure}[H]
      		\centering
            \includegraphics[width=0.8\linewidth]{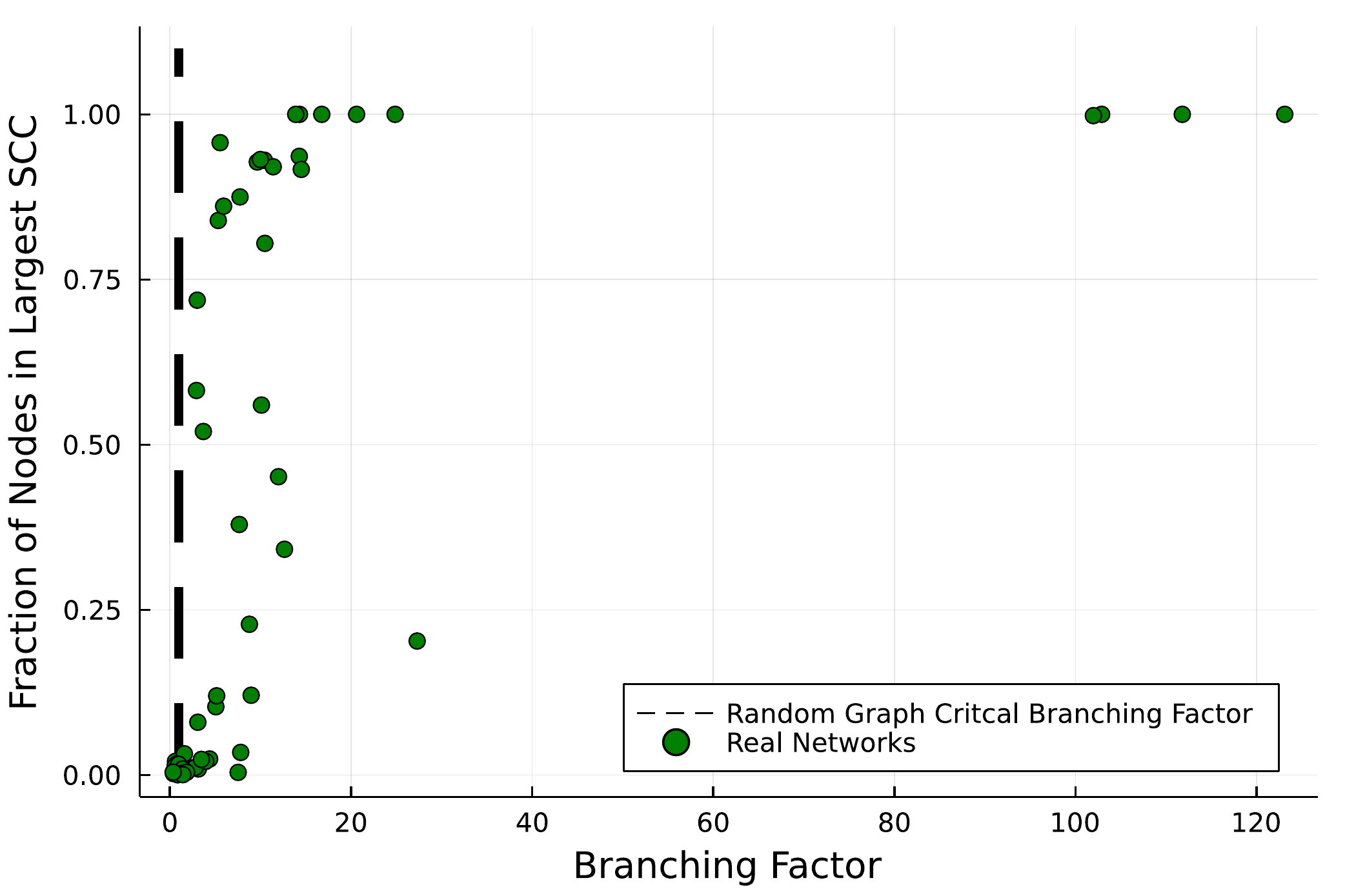}
        	\caption{ Prediction of Strong Connectivity using the Branching Factor for Real Networks \cite{DataSamJohnson}.    
        	}
        	
        \label{fig:Connectivty_branching_all}
        \end{figure}

The branching factor is shown for all the real networks in figure \ref{fig:Connectivty_branching_all}. Again it is clear the the transition is missed when not considering the hierarchy and methods derived from random graphs \cite{Boguna2005GeneralizedNetworks} are not enough to explain the behaviour of real networks. 

\end{document}